\documentclass[11pt,a4paper]{article}
\usepackage[T1]{fontenc}
\usepackage[utf8]{inputenc}
\usepackage{amsmath, amsthm, amsfonts, amssymb,adjustbox}

\usepackage[a4paper,
            bindingoffset=0.2in,
            left=1in,
            right=1in,
            top=1in,
            bottom=1in,
            footskip=.25in]{geometry}

\usepackage{xcolor}
\usepackage{multirow}
\usepackage{graphicx, color}
\usepackage{xcolor}
\usepackage{multirow}
\usepackage{verbatim}
\usepackage{comment}
\usepackage{achemso}

\author{\bf Sonia Cambiaso$^{\flat\kern-1.4pt\flat,1}$,  Fabio Rasera$^{\flat\kern-1.4pt\flat ,2}$, \and Antonio Tinti$^2$, Davide Bochicchio$^1$, Yaroslav Grosu$^{3,4}$, \\  Giulia Rossi$^{1,\bigstar}$, Alberto Giacomello$^{2,\bigstar}$}
\date{%
    $\flat\kern-1.4pt\flat$ These authors contributed equally \\[1ex]
	$^1$Dipartimento di Fisica, Universit\`a di Genova, 16146 Genoa, Italy\\[1ex]
    $^2$Dipartimento di Ingegneria Meccanica e Aerospaziale, Sapienza Universit\`a di Roma, 00184 Rome, Italy
     \\[1ex]
    $^3$Centre for Cooperative Research on Alternative Energies (CIC energiGUNE), Basque Research and Technology Alliance (BRTA), Alava Technology Park, Albert Einstein 48, 01510, Vitoria-Gasteiz, Spain\\[1ex]
    $^4$Institute of Chemistry, University of Silesia, Szkolna 9, 40-006, Katowice, Poland\\[1ex]
   $\bigstar$ \texttt{giulia.rossi@unige.it \quad alberto.giacomello@uniroma1.it}
}

\title{\bf  Local grafting heterogeneities control water intrusion and extrusion in nanopores}

\begin{document}
\maketitle

\begin{abstract}
Hydrophobic nanoporous materials can only be intruded by water forcibly, typically increasing pressure. For some materials, water extrudes when the pressure is lowered again. Controlling intrusion/extrusion hysteresis is central in technological applications, including energy materials, high performance liquid chromatography, and liquid porosimetry, but its molecular determinants are still elusive. Here, we consider water intrusion/extrusion in mesoporous materials grafted with hydrophobic chains, showing that intrusion/extrusion is ruled by microscopic heterogeneities in the grafting. For example, intrusion/extrusion pressures can vary more than 60~MPa depending on the chain length and grafting density. Coarse-grained molecular dynamics simulations reveal that local changes in radius and contact angle produced by grafting heterogeneities can pin the water interface during intrusion or facilitate vapor bubble nucleation in extrusion.
These microscopic insights can directly impact the design of energy materials and chromatography columns, as well as the interpretation of porosimetry results.
\end{abstract}

\section*{Introduction}

Hydrophobic nanoporous materials combine ``water fear'' with confined spaces, which together strongly repel water from the pores. In these conditions,  water intrudes the nanopores only if an external action is exerted, e.g., under pressure; once the intrusion process is completed, vapor bubbles must nucleate in order to give rise to the opposite phenomenon of extrusion, which typically occurs at lower pressures.
These processes can be exploited in a number of applications, which range all the way between energy storage and damping~\cite{eroshenko2001energetics, grosu2014water} , depending on the hysteresis between intrusion and extrusion pressures. 
There are other contexts in which water extrusion is undesired: in high performance liquid chromatography (HPLC), dewetting is at the origin of retention losses~\cite{gritti2019kinetic}, which makes the stationary phase unavailable to analytes, while in nanopore sensing it might cause undesirable noise in the ion current~\cite{smeets2006salt} . Finally, intrusion isotherms are used in water porosimetry~\cite{fadeev1997study,fraux2017forced} to measure the pore size distribution, interpreting them in terms of the classical Laplace's equation~\cite{rouquerol1994recommendations} . In all these cases, a connection between the microscopic characteristics of the pores (radius, contact angle, local heterogeneities, etc.) and the macroscopic properties of the system (intrusion and extrusion pressures, hysteresis, dissipated energy, etc.) is highly needed to design specific applications, control undesired phenomena, and interpret experimental results.

Arguably the most common and consolidated class of nanoporous hydrophobic materials is mesoporous silica functionalized (``grafted'') with hydrophobic organic chains, such as silanes~\cite{gauthier1996study,he2013surface,dugas2003surface} . For example, popular reverse-phase HPLC columns adopt grafted mesoporous silica gels as the stationary phase~\cite{colin1977introduction,cruz1997chromatographic,cabrera2004applications}, while similar materials are used also for vibration dampers and shock absorbers~\cite{suciu2003investigation}. These materials are inexpensive, fall in an interesting pore size range (2 to 50 nm), and boast superior mechanical stability~\cite{suciu2003investigation,grosu2015stability}. Theoretical advances in understanding intrusion and extrusion in mesoporous materials are continuously made~\cite{lefevre2004intrusion,guillemot2012,tinti2017intrusion,tinti2018,deroche2019reminiscent,ledonne2022,paulo2023impact} .

Nevertheless, the reference theoretical framework relies on the definition of average physical parameters describing the geometry of the pores and the hydrophobicity of the grafted surface, such as average radius and contact angle. A fundamental question still awaits an answer: how do the microscopic heterogeneities in the grafted pore surface affect global properties of interest for applications?

Here, we employ coarse-grained molecular dynamics simulations to reveal the {microscopic} mechanisms of water intrusion and extrusion in silanized mesoporous materials. Interestingly, even though the substrate is the same, small differences in the grafting -- different chain lengths and grafting densities -- cause significant changes in the intrusion and extrusion processes, affecting crucial technological parameters, such as the intrusion and extrusion pressures, the shape of the pressure-volume isotherms, and the dissipated energy. 
Simulations allowed us to reveal the microscopic origin of such variability, which is rooted in the local properties arising from chemical and topographical heterogeneity within the pores. 
For intrusion, the grafted layer's thickness is insufficient to justify the increase in the intrusion pressure via Laplace's equation; only by including local changes in the pore radius and in the hydrophobicity is it possible to account quantitatively for its variability.
For extrusion, on the other hand, simulations show that local accumulations of hydrophobic material {and constrictions} act as nucleation seeds for vapor bubbles. These effects were not accessible by previous simulation efforts on idealized hydrophobic nanopores without explicit functionalization~\cite{tinti2017intrusion,tinti2021} .

The present results shed light on the multiscale nature of intrusion and extrusion phenomena, revealing how molecular details control the macroscopic behavior of nanoporous materials, determining their technological applicability. In particular, physical and chemical heterogeneities in the functionalization of mesoporous materials are shown to play a crucial role, being able to change the experimental intrusion/extrusion pressures by up to 60~MPa and to almost double the dissipated energy. Controlling grafting can thus be favorably exploited to rationally design materials for various applications including energy storage, vibration damping, and liquid chromatography. In addition, microscopic insights facilitate the interpretation of liquid porosimetry measurements, revealing significant deviations from the expected macroscopic behavior due to grafting heterogeneities.

\section*{Results}
\label{sec:res}

\subsection*{Coarse-grained model of grafted mesopores}

\begin{figure}[!ht]
\centering
  \includegraphics[width=.9\linewidth]{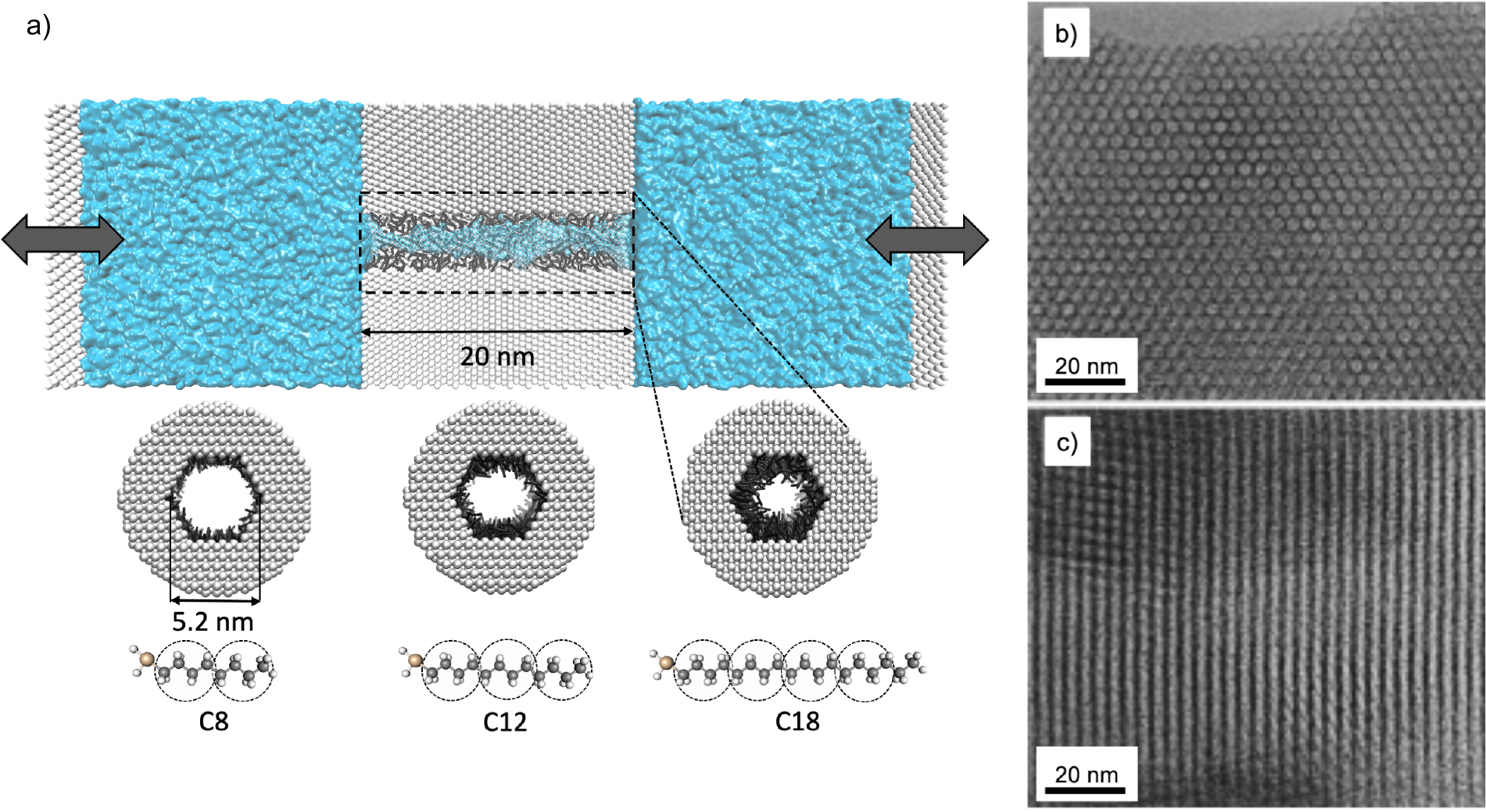}
  \caption{Setup of simulation system and TEM micrographs of MCM-41 material. a) Cross-section of the simulated system, which consists of a cylindrical pore confined between two water reservoirs whose pressure is controlled by movable pistons. The inset below shows the front view of the pore with three graftings, obtained by hydrophobic chains of different lengths (C8, C12, and C18). The molecular structures of these silanes are shown along with their respective coarse-grained scheme (Si in yellow, C in black, and H in white). 
  TEM micrographs of MCM-41 from  Liu et al.~\cite{liu2000tem} are reported in panel b (section perpendicular to the pores' axis) and c (parallel).}
  \label{system}
\end{figure}

Our goal is to build a sufficiently detailed model of a mesopore grafted with different hydrophobic chains yet with a low enough computational cost to simulate experimentally relevant time and length scales.  
To this aim, we constructed a coarse-grained (CG) model of a silica nanopore functionalized with a hydrophobic silane grafting~\cite{lefevre2004intrusion}. 
The simulated systems have geometrical similarities with the nanoporous material MCM-41, made of essentially independent, parallel cylinders with diameters ranging from 2 to 11 nm~\cite{lefevre2004intrusion,bhattacharyya2006recent} .
We used the most recent MARTINI force field (MARTINI 3~\cite{souza2021MARTINI}) in which, on average, two to four heavy atoms and associated hydrogens are mapped into one CG bead. This framework enables the computation of systems larger than those accessible to all-atom models and for longer times, with an overall speedup of at least two orders of magnitude~\cite{dejong2016MARTINI} , while maintaining sufficient chemical and spatial resolution.

We modeled the silica matrix with a slightly hydrophilic cubic fcc lattice, from which a cylindrical pore of diameter 5.2 nm and length 20 nm was excavated. The pore was immersed in water, and the liquid pressure was controlled mechanically by the force acting on two pistons far away from the pore~\cite{marchio2018} , allowing to perform in silico intrusion/extrusion (I/E) cycles analogous to experimental ones. Periodic boundary conditions were applied in the directions orthogonal to the pore axis.

To functionalize the pore, we used three alkyl (CH$_3$)C$_m$H$_{2m+1}$ linear chains having different lengths: C8 ($m=8$), C12 ($m=12$), and C18 ($m=18$). The chains were grafted to the surface at one end at random locations and oriented perpendicularly to the normal vector of the local internal surface (Supplementary Figure~S1). Each bead on the solid surface was bonded to at most one silane. For each of the three chain lengths (C8, C12, C18), we produced systems with a variety of grafting densities spanning the range $0.6$ to $1.2$~groups per square nanometer (gps nm$^{-2}$). In the following section, we characterize the degree of hydrophobicity achieved by different graftings. We remark that, albeit some features of MARTINI water do not precisely match those of actual water, the phenomenology investigated here is related to solvophobic phenomena, which are expected to be independent of the particular liquid~\cite{evans2015} .

\subsection*{Hydrophobicity of grafted surfaces}
    \begin{figure}[!ht]
    \centering
      \includegraphics[width=.9\linewidth]{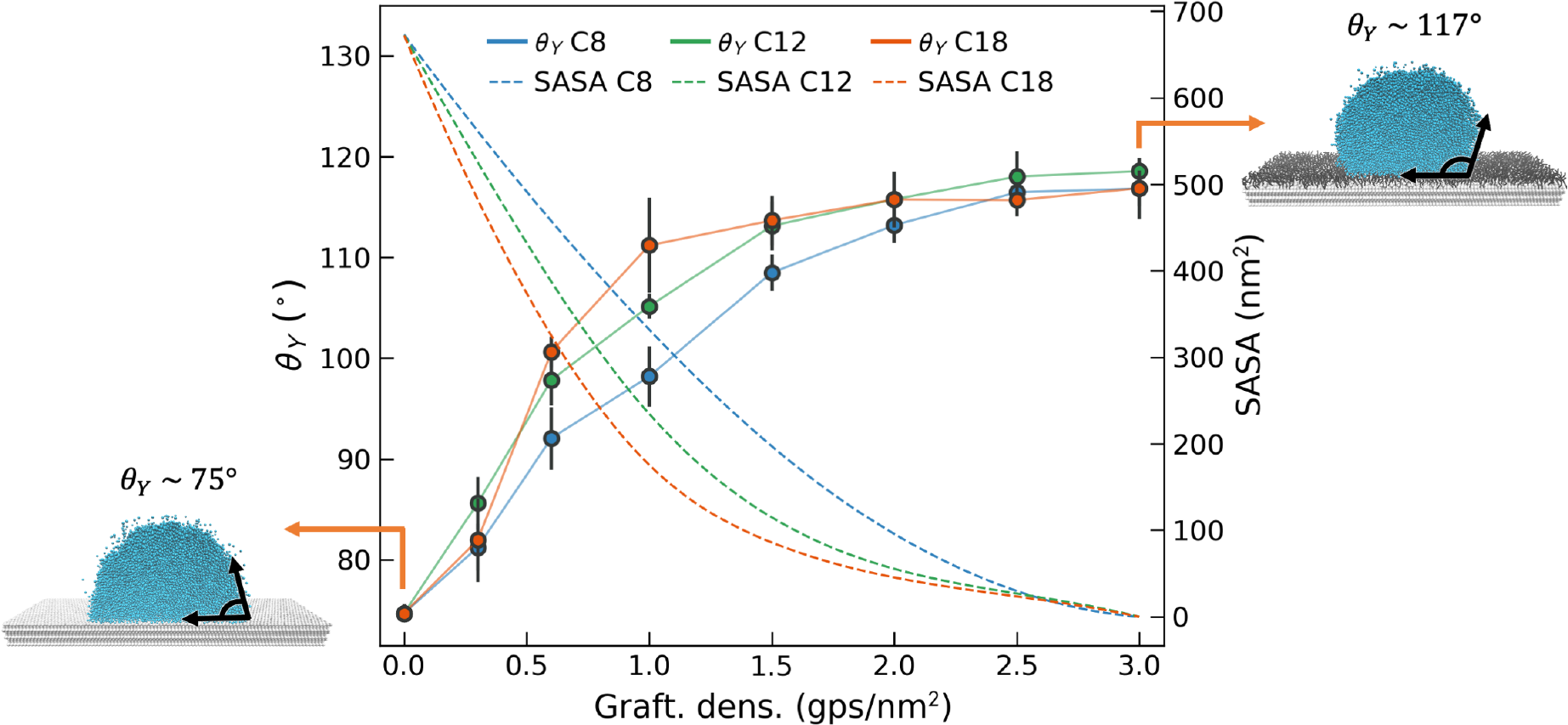}
      \caption{Contact angle $\theta_Y$ computed on a flat surface with different grafting densities and chain lengths. Each point results from an average of 5 independent simulations of cylindrical droplets deposited on the surface; error bars denote the standard deviation. A linear interpolation between data points (solid lines) is shown as a guide to the eye. The dashed lines are the average SASA of the silica surface, computed using a test particle with the same van der Waals radius as the smallest water bead. Blue, green, and orange represent C8, C12, and C18, respectively.}
      \label{CA_SASA}
    \end{figure}
    
In order to assess the effect of the grafting on the hydrophobicity of the pore, we computed the contact angle $\theta_Y$ of cylindrical water droplets deposited on a flat silica surface grafted with different chain lengths (C8, C12, and C18) within a broad range of grafting densities (0 to 3.0~gps nm$^{-2}$). The average contact angles are reported in Figure~\ref{CA_SASA}  as a function of the grafting density. 

In accordance with the experimental wisdom of silica being hydrophilic at ambient conditions{~\cite{iglauer2014contamination}} , we modeled the substrate to obtain a Young contact angle $\theta_Y\approx 75^\circ$. {Although the actual contact angle of silica has been reported within a wide range of values, its dependency with the percentage of accessible silanol groups is well known~\cite{kostakis2006effect} . For a 50\% fraction of accessible silanol groups, which would correspond to an MCM-41 material~\cite{ide2013quantification} , the contact angle is in the vicinity of $75^\circ$. For completeness, in the Supplementary section "Effect of silica hydrophobicity" we discuss I/E when more hydrophilic substrates are considered.}
Figure~\ref{CA_SASA} shows that, for grafting densities below 0.5 gps nm$^{-2}$, the interaction of water with the hydrophilic substrate dominates, showing a progressive increase of $\theta_Y$ with the grafting density; effects of the chain length are hardly discernible in this regime. At intermediate grafting densities, between 0.5 and 1.5 gps nm$^{-2}$, longer chains result in higher $\theta_Y$ for the same grafting density. Finally, at large grafting densities, the contact angle reaches a plateau at $\theta_Y\approx 117^\circ$, showing that the effect of additional chains on the hydrophobicity saturates. 
The grafted pores we used in the following sections have grafting densities 0.6, 0.8, 1.0, and 1.2~gps nm$^{-2}$, which correspond to somewhat higher effective densities because of curvature.

We further computed the Solvent Accessible Surface Area (SASA) of the silica substrate for the different graftings (dashed lines in Figure~\ref{CA_SASA}). This analysis clearly shows that the grafting diminishes the contact between water and the hydrophilic substrate, causing a progressive increase in the hydrophobicity of the grafted surface. Longer chains provide better coverage for the same grafting density, which corresponds to lower SASA; this affects the contact angle more prominently at intermediate grafting densities. As the grafting density further increases, the chains become more packed and the hydrophilic patches vanish altogether. Accordingly, the grafted layer approaches a complete coverage and the hydrophobicity becomes independent of the chain length.

\subsection*{In silico intrusion/extrusion experiments}

    \begin{figure}[!ht]
    \centering
      \includegraphics[width=1.0\linewidth]{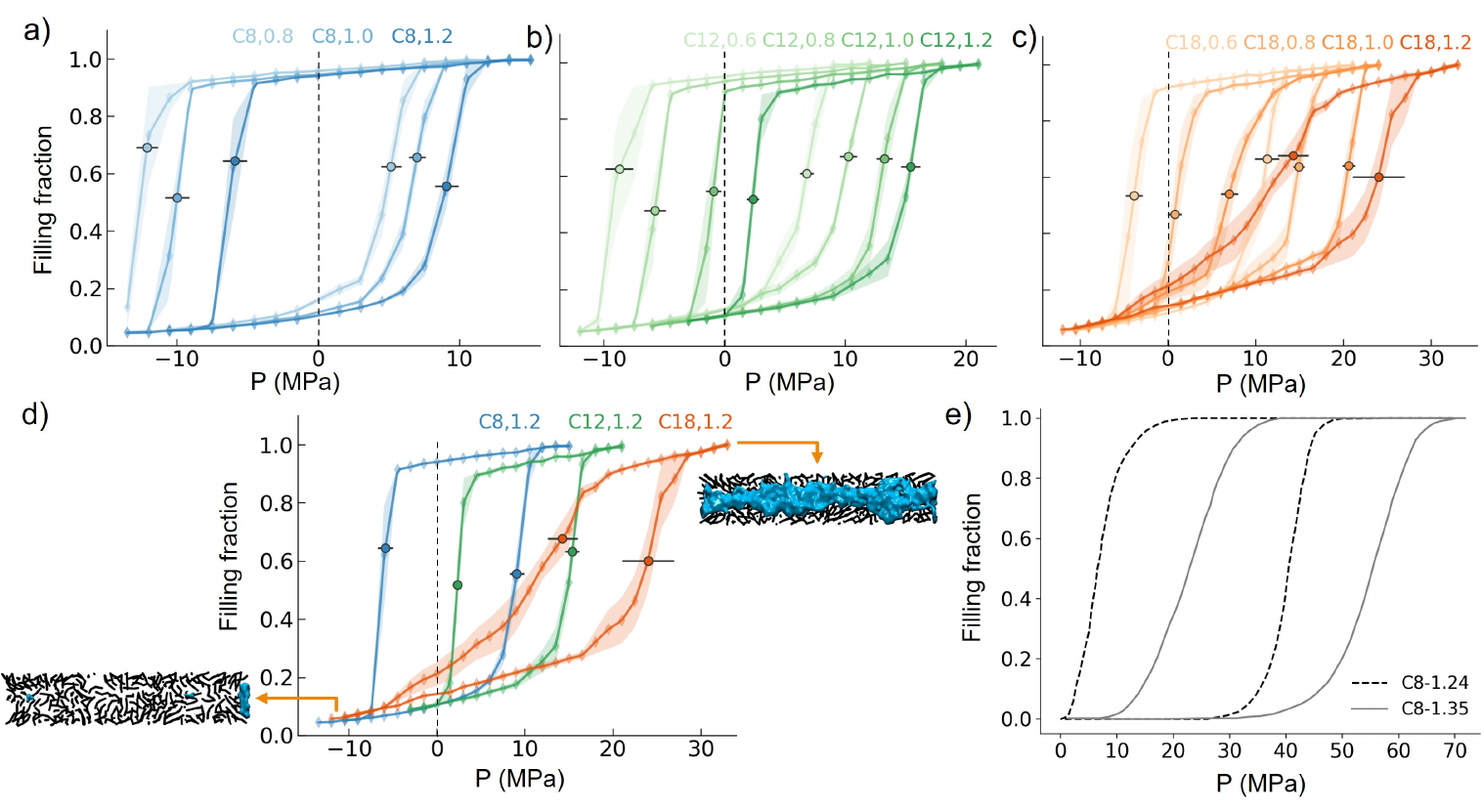}
      \caption{In silico intrusion/extrusion cycles.
      The effect of different grafting densities is shown for C8 (a), C12 (b), and C18 (c). d) I/E cycles at grafting density 1.2 gps nm$^{-2}$ for the three chain lengths; insets show the dry and wet states of the pore. These cycles mimic the typical I/E experiments that can be performed on similar materials\cite{lefevre2004intrusion}, except that, in simulations, negative pressures can be reached. e) Experimental I/E curves~\cite{lefevre2002dissipative} for an MCM-41 matrix grafted with C8 at two different grafting densities: 1.24 (dashed line) and 1.35 gps nm$^{-2}$ (solid line). We remark that experimental materials have a much smaller radius than in our simulations, $R_{C8-1.24}=1.75$~nm and $R_{C8-1.35}=2.1$~nm before grafting, and that the compressibility of the system was subtracted by the authors from the I/E curves.}
      \label{int_ext_cycles}
    \end{figure}

We performed in silico I/E cycles for all grafted pores by progressively increasing the water pressure until complete filling of the pore was achieved and subsequently decreasing it until emptying. These cycles are analogous to typical I/E experiments performed on similar materials~\cite{lefevre2004intrusion}, except that, in simulations, negative pressures can be reached, and the cycles are necessarily faster, although this is partially compensated by the shorter pores. Figure~\ref{int_ext_cycles} illustrates the I/E curves obtained for different grafting densities (panels a-c) and chain lengths (d). The most apparent finding is that there are significant qualitative and quantitative changes in the intrusion and extrusion curves for a relatively narrow change in chain lengths (C8-C18) and grafting densities (0.6-1.2~gps nm$^{-2}$). Intrusion and extrusion pressures vary by $>20$~MPa and the shapes can account for rather abrupt I/E processes, as expected for monodisperse cylindrical pores, to rather progressive ones (C18, 1.2~gps nm$^{-2}$).

Figure~\ref{int_ext_cycles}a-c shows that higher grafting densities shift both the intrusion and the extrusion curves to higher pressures, for all considered chain lengths. Panel d reveals that, for the same grafting density, the I/E curves move to higher pressures as the chain length is increased.
Our data show the same qualitative behavior -- growth of the I/E pressures with grafting density-- as the experimental data (Figure~\ref{int_ext_cycles}e) for an MCM-41 matrix with different grafting densities and radii~\cite{lefevre2002dissipative} . It is also worth noticing that our simulations capture the change in slope with grafting density observed in such experiments.

    \begin{figure}[!ht]
    \centering
      \includegraphics[width=.8\linewidth]{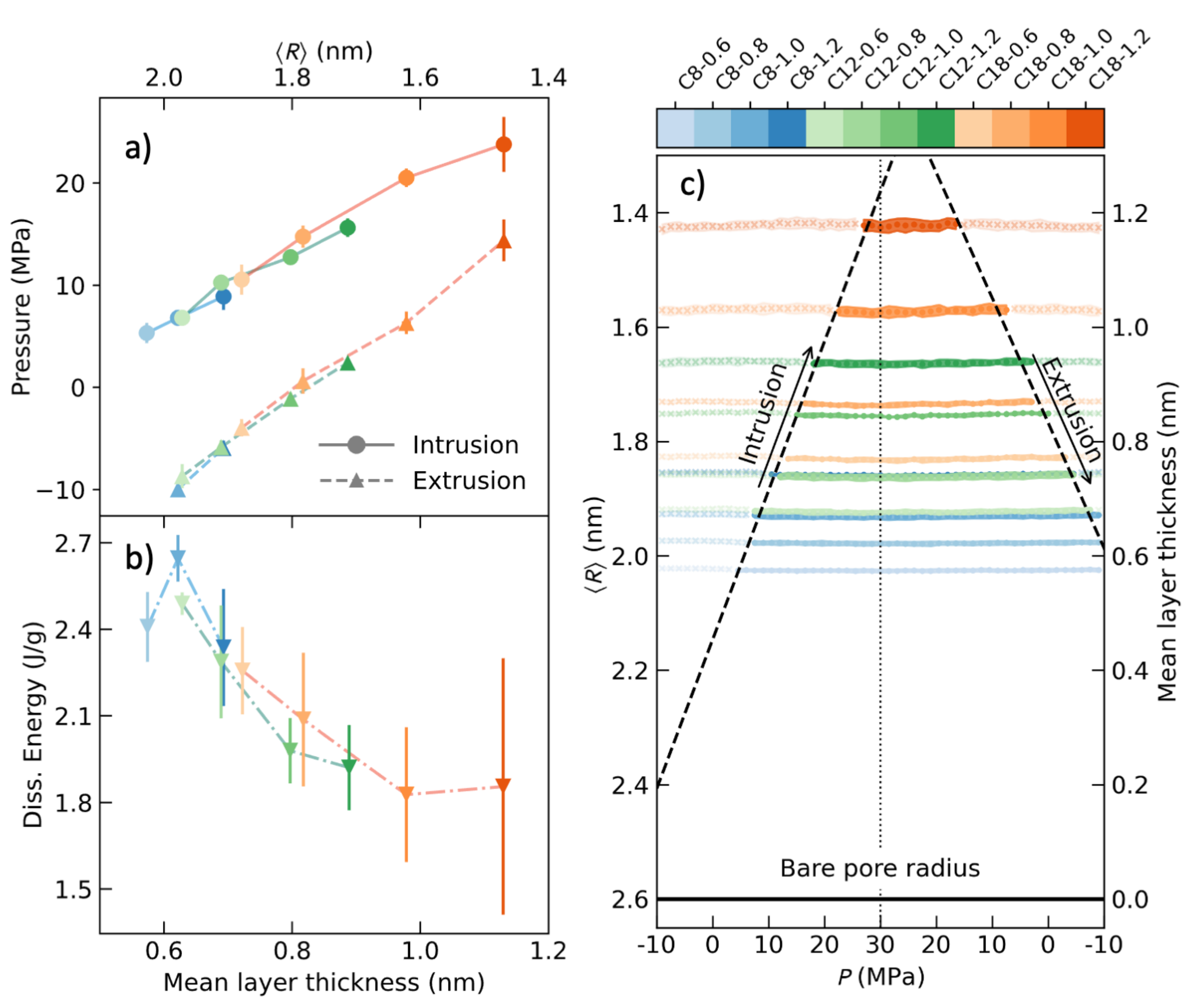}
      \caption{On the left, relation between intrusion/extrusion pressures and dissipated energy with $\langle R \rangle$ (mean layer thickness). On the right, $\langle R \rangle$ (mean layer thickness) variation with pressure. a) I/E pressures, defined as the inflection points of the respective curves, against mean layer thickness and $\langle R \rangle$. b) Dissipated energies for each system, as computed by multiplying the area between extrusion and intrusion curves times the typical pore density of MCM-41~\cite{guillemot2012new} (330 mm$^3$/g ). 
      c) Mean layer thickness and mean radius of the grafted pores as a function of pressure. Notice that the pressure increases until 30 MPa, shown with a vertical line, then starts decreasing. $\langle R \rangle$ is calculated as $R_\mathrm{pore}$ subtracted of the mean layer thickness; to calculate distances, we use the point position of the beads, in consistency with that used for computing the contact angles in Figure~\ref{CA_SASA}. The cross/circular markers represent dry/wet configurations, delimited by a linear fit to the I/E pressures. The color legend refers to both figures. {More information on how to compute the mean radius, mean layer thickness, and dissipated energy can be found in the "Supplementary Method" section in SI.}}
      \label{Radius_dist}
    \end{figure}

In the attempt to rationalize the effect of grafting on I/E, we note that both the chain length and the grafting density are expected to affect the size of the pore available to the water. Indeed, Figure~\ref{Radius_dist}c shows that the mean thickness of the grafting layer grows with chain length and grafting density, modifying the mean radius of the grafted pore $\langle R \rangle$ between ca. $1.4$ and $2$~nm. The variance in $\langle R \rangle$ increases with grafting density, with the maximum attained for C18.  
The compression of the grafted layer upon intrusion is seen to be negligible: the softness of the ligands does not seem to play a critical role for the considered systems. In fact, Supplementary Figure~S6 shows that the mean radius increases/decreases no more than 0.02 nm during intrusion/extrusion. {In addition, performing I/E cycles with artificially immobilized silanes does not alter the results, as shown in Supplementary Figure~S8.}

The difference in $\langle R \rangle$ across systems suggests that the amount of hydrophobic material added inside the hydrophilic channel is key to understanding the I/E mechanisms. Figure~\ref{Radius_dist}a indeed highlights that the intrusion and extrusion pressures $P_\mathrm{int/ext}$ grow monotonically with $\langle R \rangle$, in agreement with experimental data~\cite{fadeev1997study} .
Similarly, the dissipated energy $E_d$ per mass of porous material depends on $\langle R \rangle$, Figure~\ref{Radius_dist}b. One can notice that $E_d$ for the lowest grafting density and shortest chain is ca. 1.5 times the energy of the highest grafting density and longest chain, consistent with previous results which reported an increase of the dissipated energy with pore radius~\cite{tinti2017intrusion} .

Overall, these results suggest that the type and quality of grafting can significantly change the macroscopic behavior of hydrophobic mesoporous materials, affecting their technological applicability. For instance, not all the considered materials display extrusion at positive pressures, which allows for a single-time use only (e.g., for ``bumpers''~\cite{eroshenko2001energetics}) . In the following, we dissect the microscopic origin of such unexpected variability, showing that {the effect of $\langle R \rangle$ considered in Figure~\ref{Radius_dist} just accounts for a general, qualitative trend, while} local variations in pore radius and hydrophobicity due to grafting heterogeneities {are the quantitative key to understand} intrusion and extrusion.

\subsection*{Microscopic origin of intrusion}
\begin{figure}[!ht]
    \centering
      \includegraphics[width=1\linewidth]{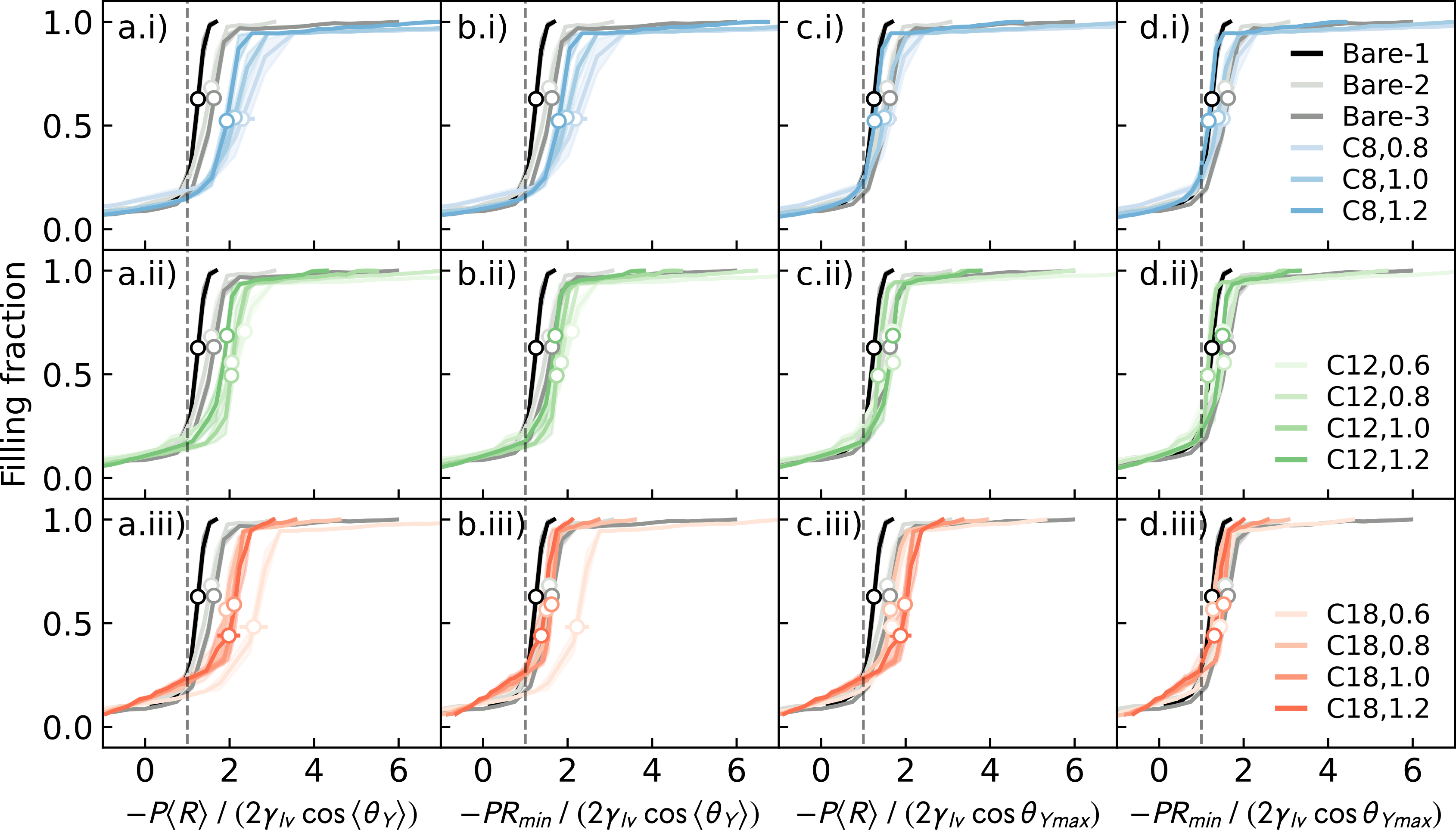}
      \caption{Intrusion curves for all the grafted and for three bare pores, i.e., without any grafting and with different substrate hydrophobicities and radii (Bare 1: $\theta_{Y}=100^\circ$ and $R=1.3 \,\textrm{nm}$, Bare 2: $\theta_{Y}=100^\circ$ and $R=1.8 \,\textrm{nm}$, and Bare 3: $\theta_{Y}=105^\circ$ and $R=1.3\, \textrm{nm}$). A summary of the grafted pore measured quantities can be found in Supplementary Table~S1. Bare nanopores details can be seen in Supplementary Figure~S9. a) Intrusion curves normalized according to Laplace's law~\eqref{Laplace} using the mean radii and contact angles.  The points show the inflection point of the curve, which we define as $P_\mathrm{int}$. The vertical line shows where $P_\mathrm{int}$ is exactly equal to Laplace's equation prediction. In b) the curves are rescaled using the minimum radii and mean contact angle, while in c) we used mean radii and maximum contact angles. Details on the calculation of the maximum contact angle can be found in the Supplementary Information. Panel d) shows the rescaling with both the local parameters. i, ii, and iii labels data refer to C8, C12, and C18 chains, respectively.}
      \label{Collapse_v2}
    \end{figure}
\begin{figure}[!ht]
    \centering
      \includegraphics[width=1.0\linewidth]{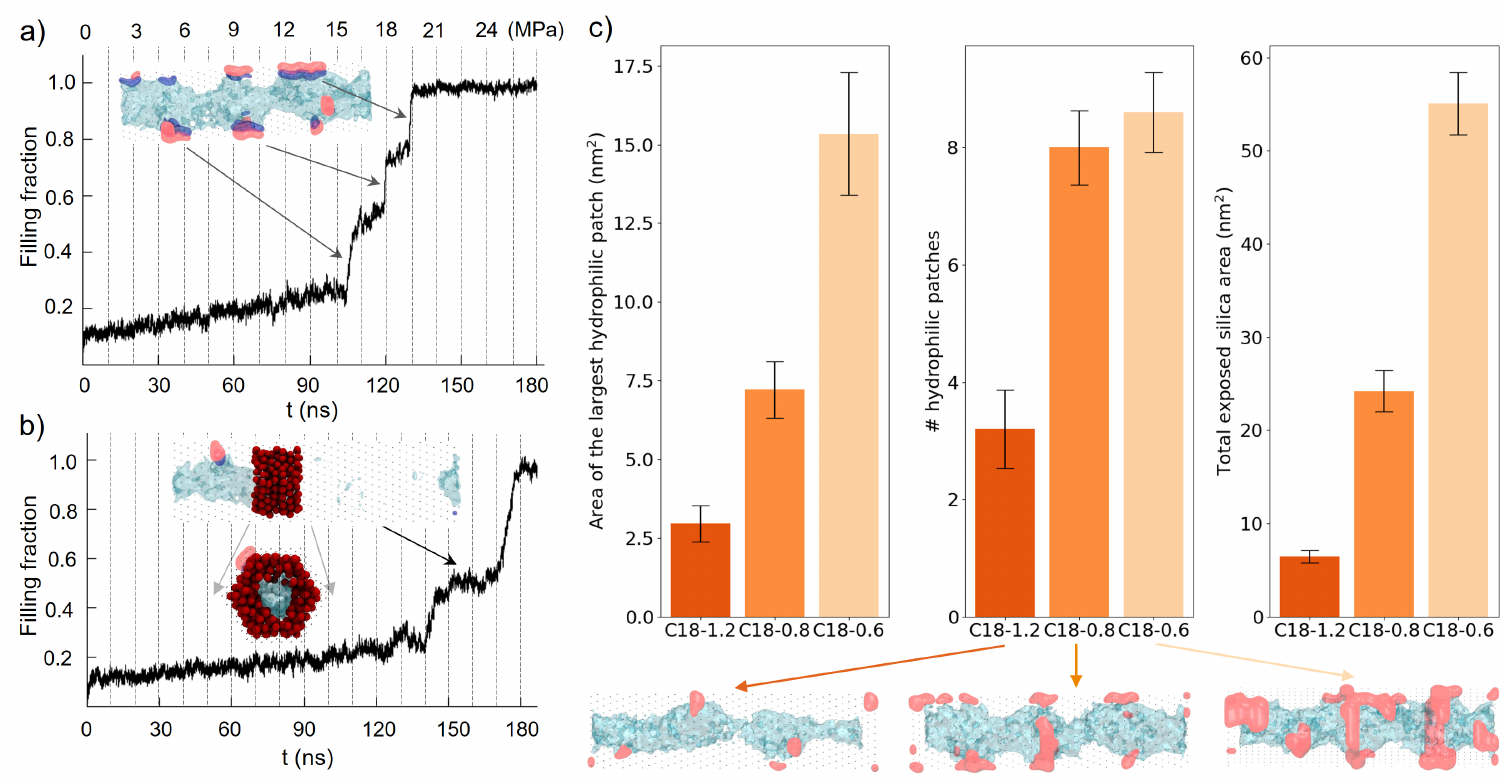}
      \caption{Details of pore heterogeneities: patches and constrictions affect water filling rates. Panels a) and b) show the water filling fraction versus time relative to single runs for C18-0.8 and C18-1.2, respectively. {Vertical lines show the times at which pressure steps are applied to the system.} The insets show water pinning inside the channel. In panel a) the hydrophilic patches are highlighted in pink; panel b) shows a local constriction, in which the radius of the pore has a minimum due to the presence of the silanes (in red).
      {c) Quantitative analysis of the hydrophilic patches in C18 systems with different graftings: the first panel shows the area of the largest hydrophilic patch, the second the total number of patches, and the third panel reports the total area of bare silica exposed. The error bars refer to the standard deviation of the values for the different realizations of each system.}
      } 
      \label{patches}
    \end{figure}
Previous results showed that the characteristics of grafting, i.e., chain length and grafting density, affect both the overall degree of hydrophobicity ($\theta_Y$, Figure~\ref{CA_SASA}) and the mean radius ($\langle R \rangle$, Figure~\ref{Radius_dist}) of the pores. Are these average quantities sufficient to explain the changes observed in the intrusion pressure across different systems? To answer this question, we employ the macroscopic Laplace equation, which describes the pressure $P_\mathrm{int}$ after which the meniscus depins from the cavity mouth, thus giving rise to intrusion~\cite{giacomello2020}:
\begin{equation}
    P_\mathrm{int} = -\frac{2\gamma_\mathrm{lv} \cos \theta_Y}{R} \ ,
    \label{Laplace}
\end{equation}
where $\gamma_\mathrm{lv}$ is the liquid/vapor surface tension and $R$ the pore radius. According to eq.~\eqref{Laplace}, the intrusion pressure should increase with higher contact angles and lower radii, which is in agreement with the {general} trends shown in Figure~\ref{int_ext_cycles}. 

For a quantitative comparison, in Figure~\ref{Collapse_v2}a.i-iii we report the intrusion curves rescaled using eq.~\eqref{Laplace} adopting the average contact angles $\theta_Y$, the mean radius $\langle R \rangle$ of all the replicas for each pore configuration (different grafting density and chain length), and $\gamma_\mathrm{lv}=21\pm1$~mN m$^{-1}$ which was calculated using the test-area method~\cite{gloor2005test}.

If only the mean values of $\theta_Y$ and $\langle R \rangle$ were relevant for the intrusion process, the rescaled curves relative to all different graftings would overlap. However, a true collapse of the curves is obtained only for the bare nanopores made of different materials and with different pore radii, while the grafted nanopores get closer to each other but do not overlap. The systems that exhibit higher deviations from the predicted behavior are those with lower amounts of grafting; this trend is unexpected since lower hydrophobicity and higher radius should result in lower intrusion pressures. Therefore, Figure~\ref{Collapse_v2}a reveals that the average radius and contact angle do not fully account for the variability of the I/E curves.

The intrusion curves in the plot are the results of an average between cycles obtained starting from different realizations of the same system, i.e., with the same macroscopic parameters (grafting density and chain length) but different microscopic configurations, with random distributions of the ligands. The intrusion curves in Figures~\ref{patches}a and b are relative to individual intrusion events of C18-0.8 and C18-1.2 systems, respectively. These curves present larger slopes than those of bare channels, as shown in Supplementary Figure~S10. The steps in the intrusion curves reflect a microscopic stick-and-slip mechanism that water undergoes during intrusion. The free-energy profile for water intrusion (Figure~S11 in Supplementary "Free energy calculations" section) indeed reveals the presence of intermediate free energy barriers between the dry and wet states. Microscopically, the pore surface is characterized by alternations of constrictions, namely local reductions of the pore radius due to grafted silanes, and hydrophilic patches, namely (sub)nanometer areas of exposed bare silica: when water meets hydrophilic patches, it is locally attracted and thus slips. When water meets constrictions, it is locally pinned and thus sticks. These microscopic heterogeneities can be quantified in terms of local variations of pore radius and contact angle.

In order to verify which of these two local parameters is the most relevant, we rescaled the intrusion curves using the minimum pore radius and the average contact angle or the maximum contact angle and the average radius of each individual trajectory, as shown in Figure~\ref{Collapse_v2}b and c, respectively. Figure~\ref{Collapse_v2}b shows that the rescaling with the minimum radius of the nanopore and the mean contact angle generally improves the agreement of the pores with higher grafting density with the bare ones. For C18, a  collapse of all curves is obtained except C18-0.6. Overall, this analysis shows that, at higher grafting densities, intrusion is affected by topographical heterogeneities, which are capable of pinning the meniscus within the pore. 

On the other hand, rescaling intrusion curves with the maximum contact angle and the mean pore radius (Figure~\ref{Collapse_v2}c) improves the curves' overlap with the bare pores for most systems except C18-1.0 and 1.2, accounting for the importance of chemical heterogeneities {(the appearance of hydrophilic patches)}, which are indeed expected at lower grafting densities and for shorter chains, due to the higher SASA of the substrate (Figure~\ref{CA_SASA}).
{More quantitatively, as shown in Figure~\ref{Collapse_v2}c, the number of hydrophilic patches for C18-0.6 and C18-0.8 is more than twice than in C18-1.2. On the other hand the typical size of hydrophilic patches, including the largest one, is larger for C18-0.6 resulting in a  total amount of hydrophilic silica exposed which is significantly higher than in the other two systems. This analysis confirms the presence of two kinds of heterogeneities (topographical and chemical),} but the distinction is not necessarily sharp, as hydrophobic chains can affect both the local radius and contact angle leading to interface pinning; this scenario seems to apply to intermediate cases, which are indeed rescaled correctly using either local quantity.
Finally, if both local quantities are taken into account, all curves tend to collapse into a master curve (Figure~\ref{Collapse_v2}d).

This analysis clarifies that the intrusion mechanism is severely dependent on local heterogeneities which need to be taken into account to predict changes to the intrusion process. For more compact graftings, topographical patchiness tends to pin the meniscus by creating bottlenecks inside the channel {(inset of Figure~\ref{patches}b, displaying C18-1.2)}. The systems with lower amounts of hydrophobic grafting are more affected by chemical patchiness, with the interplay of hydrophobic and hydrophilic areas controlling water intrusion (inset of Figure~\ref{patches}b, showing C18-0.8).
To complete wetting in grafted pores, at variance with bare pores, water must overcome the most severe pinning conditions, which can be topographical (minimal radius),  chemical (most hydrophobic patch), or combinations thereof~\cite{giacomello2016} .

\subsection*{Microscopic origin of extrusion} \label{ssec:extrusion}
    \begin{figure}[!ht]
    \centering
      \includegraphics[width=1\linewidth]{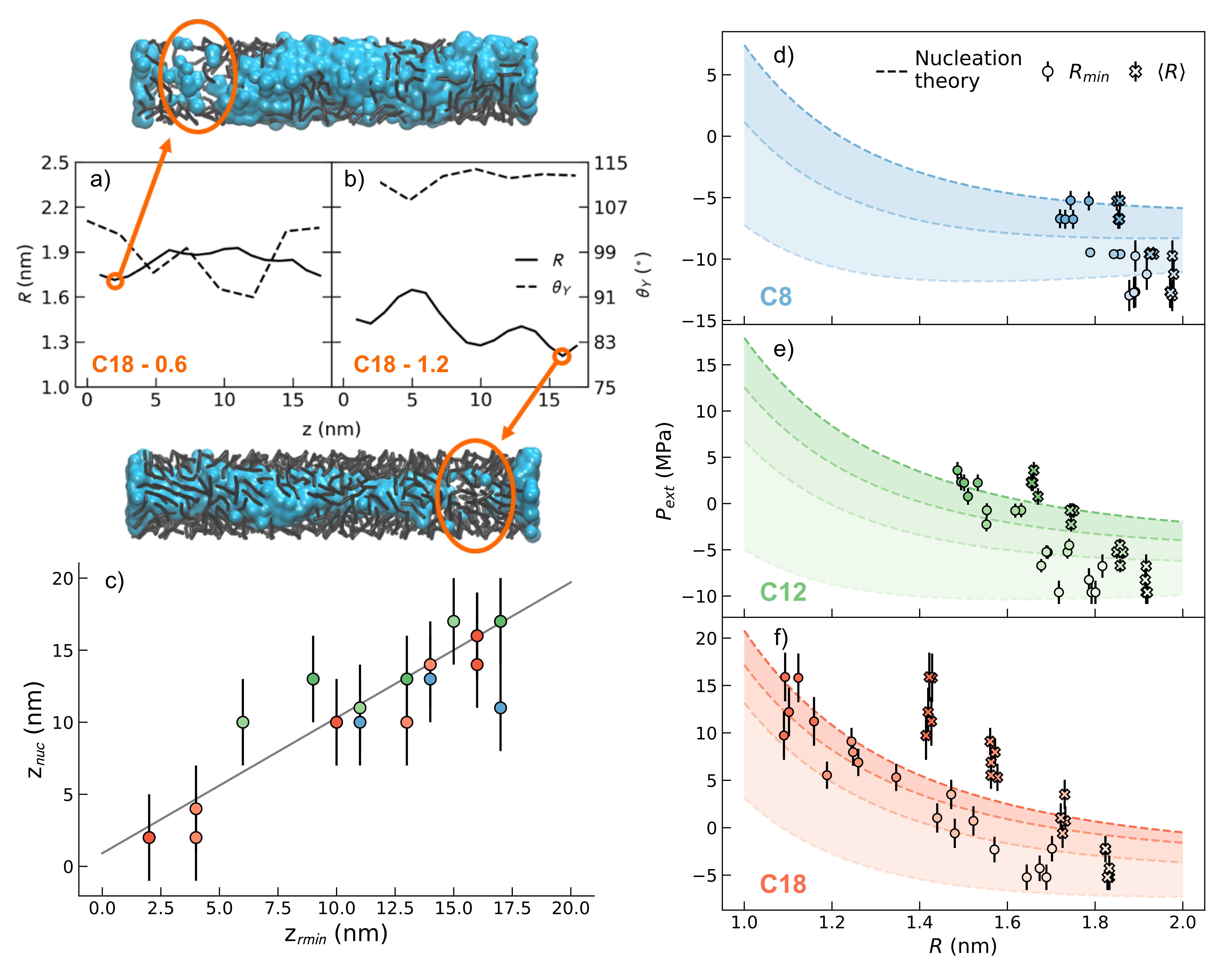}
      \caption{Local variations of radius and contact angle determine preferential nucleation sites. Nucleation theory can better describe simulation results when minimum radii are provided. The local radius of the grafted pore and local contact angle $\theta_Y$ is shown as a function of the axial coordinate for two representative systems with a) low (C18-0.6) and b) high (C18-1.2) grafting densities. The match between high contact angle and small radii favors nucleation, as shown by the critical bubbles in the reported MD snapshots. c) Position of the nucleation bubble along the pore ($z_\mathrm{nuc}$) as a function of the position of the minimum radius along the pore ($z_\mathrm{min}$). d-f) Extrusion pressure $P_\mathrm{ext}$ for individual MD simulations as a function of the minimal radius $R_\mathrm{min}$ (filled circles) and of the average radius $\langle R\rangle$ (filled x). The predictions of the nucleation theory~\eqref{nucleation_theory} are reported in dashed lines for the average contact angles $\theta_Y$ of each grafting density and chain length combination. The shaded areas are used to demarcate the limiting curves for each chain length: blue for C8, green for C12 and red for C18.
      }
      \label{Extrusion}
    \end{figure}

Although it can be described within the same framework~\cite{tinti2017intrusion}, the conditions at which extrusion occurs are not described by Laplace's equation~\eqref{Laplace}. Actually, extrusion originates in the nucleation of a vapor phase from the confined liquid (cavitation), which involves the nontrivial competition of bulk, surface~\cite{kelton2010nucleation,giacomello2020}, and line~\cite{lefevre2004intrusion,guillemot2012,tinti2017intrusion} terms. Thus, thermal fluctuations can stochastically lead to the emptying of an individual pore by overcoming a free energy barrier $\Delta \Omega^\dag$. 

By looking at individual extrusion trajectories, we noticed that extrusion starts at specific locations within the pore characterized by the smallest radius (Figure~\ref{Extrusion}a and b, Supplementary Figure~S12-S13). {Figure~\ref{Extrusion}c shows a fair correlation between the position of the nucleation bubble along the channel and the position of the minimum radius.} This region usually also corresponds to the most hydrophobic region, because at those locations more hydrophobic material is accumulated. {If along the pore there are multiple constrictions, multiple bubbles can nucleate at the same time, as shown in Supplementary Figure~S14.} Together, constrictions and enhanced hydrophobicity, which are maximum at the highest grafting densities, conspire in favoring the formation of a cavitation nucleus. 

More formally, expressing the nucleation free energy as a sum of volume, surface, and line terms, one can build a simple classical nucleation theory for the formation of a bubble in a cylindrical pore~\cite{lefevre2004intrusion,tinti2023} . In particular, the nucleation barrier $\Delta \Omega^\dag$ can be expressed as 
\begin{equation}
    \Delta\Omega^{\dag} = \Delta P V^* + \gamma_\mathrm{lv} (A^*_\mathrm{lv} +\cos\theta_Y A_\mathrm{sv}^*)+ \tau l^*_\mathrm{slv}
    \label{nucleation_theory}
\end{equation}
where $V^*$, $A^*_\mathrm{lv}$, $A^*_\mathrm{sv}$, and $l^*_\mathrm{slv}$ are the volume, liquid-vapor area, solid-vapor area, and triple line length of the critical bubble, respectively; $\tau$ is an effective line tension that takes into account nanoscale effects, see, e.g., \cite{tinti2017intrusion, schimmele2007conceptual, schimmele2009line} . We used the estimates of Lefevre et al.~\cite{lefevre2004intrusion} for the geometrical quantities. For the thermodynamic parameters, we kept a constant value of $\Delta \Omega^\dag=10$~k$_\mathrm{B}$T, which is compatible with the simulated times, and considered a range of $\tau$ close to the ones previously reported in literature~\cite{tinti2017intrusion, paulo2023,badr2022cloaking} , $-10$~pN $<\tau<-5$~pN). 

Figures~\ref{Extrusion}d-f compare $P_\mathrm{ext}$ of individual simulations with the nucleation theory~\eqref{nucleation_theory}.
We use a definition of $P_\mathrm{ext}$ which is slightly different from Figure~\ref{Radius_dist}: we take the point in which the water volume starts a steep decrease instead of using the inflection point of the extrusion curve. In this way, we select with more precision the pressure at which a vapor phase nucleation starts, which is the process described by the theory. We further consider the minimal radius $R_\mathrm{min}$ of each pore (local quantity) and the average contact angle for each chain length/grafting density pair.
With these quantities, the theory seems to capture well the trend of $P_\mathrm{ext}$, showing that higher values are connected to larger $\theta_Y$. In contrast, the average value $\langle R \rangle$ predicts extrusion pressures which are far off, especially for C18.

It is worth remarking that both $\tau$ and $\Delta \Omega^\dag$ are expected to vary for each individual system, reflecting the variability in the grafting. Hence, a more precise quantitative agreement could be expected only if one had a careful measurement of these quantities and a nucleation theory that takes into account the non-ideal geometry of the pore. 

In summary, for extrusion, heterogeneities in the grafting gives rise to variability in the local radius and hydrophobicity that determine where the most probable sites for nucleation are located. The extrusion pressures can be estimated using a nucleation theory, which however needs to be informed with local quantities that are hard to characterize $\tau$, $\Delta \Omega^\dag$, and $R_\mathrm{min}$, in addition to $\theta_Y$ of the specific grafting.

Our Molecular Dynamics results,  demonstrating the pivotal importance of the molecular features of the grafting on the I/E process, were found to be in agreement with recent numerical simulations\cite{doebele_thesis} which were performed by means of a minimal lattice model, along with specialized experiments \cite{picard2021dynamics} which were conducted on straight pores functionalized with extremely controlled and reproducible grafting patterns.

\section*{Discussion}
\label{sec:dis}

Before discussing the impact of our work on three different technological applications, it is worth discussing in more detail the possibility to achieve a quantitative comparison between the simulated I/E cycles and the experimental ones. There are three main sources of discrepancies between simulated and experimental data which need to be considered: pore geometry, forcefield used in simulations and time and length scale differences between experiments and simulations.

The experimental samples obviously exhibit a more heterogeneous pore geometry, including pore interconnections, structural defects, and polydispersity in lengths and radii, which are not captured by the simple cylindrical pore geometry used in our simulations. These effects, whose understanding is still in its infancy \cite{amabili2019} , will add up to the effect of heterogeneities in the grafting which is the focus of this work.

The I/E process is ruled by interfacial energies. Atomistic forcefields can reasonably approximate experimental surface and line tensions. Coarse grained models, and the MARTINI forcefield in particular, have the tendency to underestimate them. While our coarse grained approach is expected to capture qualitatively the physico-chemical driving forces for I/E processes, a comparison between experimental and simulated intrusion pressures requires the application of a correction factor. As the liquid-vapor surface tension of water in the MARTINI forcefield is calculated as $\gamma_\mathrm{sim}=21$ mN m$^{-1}$ while the experimental value is $\gamma_\mathrm{exp}=71.7$ mN m$^{-1}$~\cite{vargaftik1983} , simulated pressures (as well as dissipated energies) should be multiplied by a factor of $\gamma_\mathrm{exp}/\gamma_\mathrm{sim}=3.41$ to be compared to the experimental ones. This rescaling  for instance predicts that the maximum variability of $P_\mathrm{int}$ and $P_\mathrm{ext}$ due to heterogeneities in the experimental case is 60~MPa. In Supplementary Figure~S15, we show a quantitative agreement between simulated and experimental \cite{fadeev1997study} intrusion and extrusion pressures as a function of grafting density, once this rescaling is taken into account.

Typical lengths of experimental MCM-41 pores, which are the archetype of our simulations, can be up to several $\mu$m long \cite{beck1992}, while typical experimental time scales are of the order of seconds \cite{guillemot2012new} . On the other hand, the simulated pore length in this work are 20~nm, while constant-pressure simulation windows had a duration of 10 ns. We have addressed in detail the quantification of the maximum discrepancy between experimental and simulated intrusion and extrusion pressures that may emerge from these different time and length scales. The details can be found in the Supplementary section ""Bridging simulations and experiments: effect of time and length scales on experimental and simulated I/E pressures", but we remark that for $P_\mathrm{int}$ the expected discrepancy is tens of kPa while for $P_\mathrm{ext}$ is of the order of few MPa, which we consider fully satisfying, given the overall goal of elucidating the physics of I/E in grafted nanopores.

We now discuss the results from the perspective of three typical technological applications of water intrusion/extrusion in grafted silica nanopores, namely energy materials, high performance liquid chromatography, and porosimetry.

I/E of liquids within hydrophobic pores is at the heart of a broad family of energy materials~\cite{ledonne2022}: depending on the I/E hysteresis, such systems can be used for vibration damping or energy storage applications. In both cases, the design for the specific operative conditions hinges on the ability to control $P_\mathrm{int}$, $P_\mathrm{ext}$, and their range. Until now, the complexity of the pores and their nanometric size hindered a satisfying characterization of the I/E behavior of promising materials, like suitably coated porous silica. Herein, our coarse-grained approach allowed us to derive important structure-property trends for energy applications, showing that grafting details cannot be ignored to control I/E cycles precisely. 

As shown in Supplementary Table~S1, in simulations I/E pressures can vary more than 20~MPa for the same bare pore radius, depending on the length and density of the functionalizing chains {(this corresponds to 68~MPa in experiments)}. For instance, using the longest ligands (C18), one can double the grafting density to more than double the intrusion pressure, moving at the same time the extrusion from negative pressures to almost 14 MPa {(48~MPa in experiments)}. Our results also suggest that small grafting densities with short ligands are to be preferred to increase the dissipated energy because these coatings preserve some hydrophilic patches. However, in these conditions, often no extrusion is possible at positive pressures, and thus the material becomes suitable for single-use energy dissipation (so-called bumpers~\cite{eroshenko2001energetics}). 

In summary, the information gathered via coarse-grained simulations (summarized in Table~S1) is vital to learn how to tune the operative conditions and the stored and dissipated energy by proper nanopore functionalization. In future studies, our coarse-grained approach will be extended to include more intricate pore geometries, accounting for the full complexity of porous materials for energy applications.

Results from our simulation campaign are of interest also for High-Performance Liquid Chromatography (HPLC), an analytical technique used to separate, identify, and quantify components in a mixture. HPLC is common in several industrial fields, such as pharmaceutical, biotech, and food processing and safety. Reversed-phase liquid chromatography (RPLC) is a type of HPLC that uses a non-polar stationary phase made of a hydrophobic material, such as C8 and C18-functionalized silica, and a polar mobile phase to analyze mixtures containing apolar and moderately polar compounds. The analysis of very polar compounds may require 100$\%$ aqueous mobile phases, which are associated with irreversible retention losses, as explained by Gritti~\cite{gritti2019kinetic} . Both experiments~\cite{Walter1997} and Monte Carlo simulations~\cite{Sun2007} revealed that dewetting of water from the hydrophobic stationary phase (also wrongly referred to as ``phase collapse'') is the reason behind retention loss, which makes part of the RPLC column unavailable to retain analytes. Several parameters can affect the dewetting kinetics: temperature, pore size, and water contact angle. However, the internal microstructure of the material, namely the pore size distribution, the surface coverage and chemistry, the pore connectivity~\cite{amabili2019,paulo2023impact} , and the presence of dissolved gases~\cite{camisasca2020} affect the retention loss as well. 
In practice, to minimize the loss of retention, the hydrostatic pressure in the column should never go below the extrusion pressure; ideally $P_\mathrm{ext}<0$. 
Large intrusion pressures, on the other hand, are indicative of the difficulty of filling the column for the first time or of regenerating one that underwent dewetting. 

The present results show that the grafting can significantly affect both the intrusion and the extrusion pressures (Supplementary Table~S1) and should be carefully considered in the design of HPLC columns. For instance, longer chain lengths and higher grafting densities are shown to exacerbate the dewetting problem and make column regeneration more difficult.
The grafting details are known to impact the selectivity towards certain analytes~\cite{gritti2020} and the grafting densities are related to the retention of analytes in chromatography~\cite{nagase2008effects} .
For this reason, our coarse-grained approach can be a predictive tool to guide the challenging research of a balance between minimal retention losses and maximum selectivity.

Finally, we proceed to discuss the importance of the present results for the field of porosimetry. Indeed intrusion experiments that are conceptually similar to those performed here in silico are routinely used in porosimetry to infer geometrical information about the pores themselves. Depending on the intruding liquid, we distinguish between water and mercury porosimetry, with the latter being a standard technique due to the very high surface tension of mercury and its typical high contact angles on most surfaces, even non-functionalized ones~\cite{schlumberger2021characterization,lefevre2004intrusion}.

\begin{figure}[!ht]
\centering
  \includegraphics[width=1\linewidth]{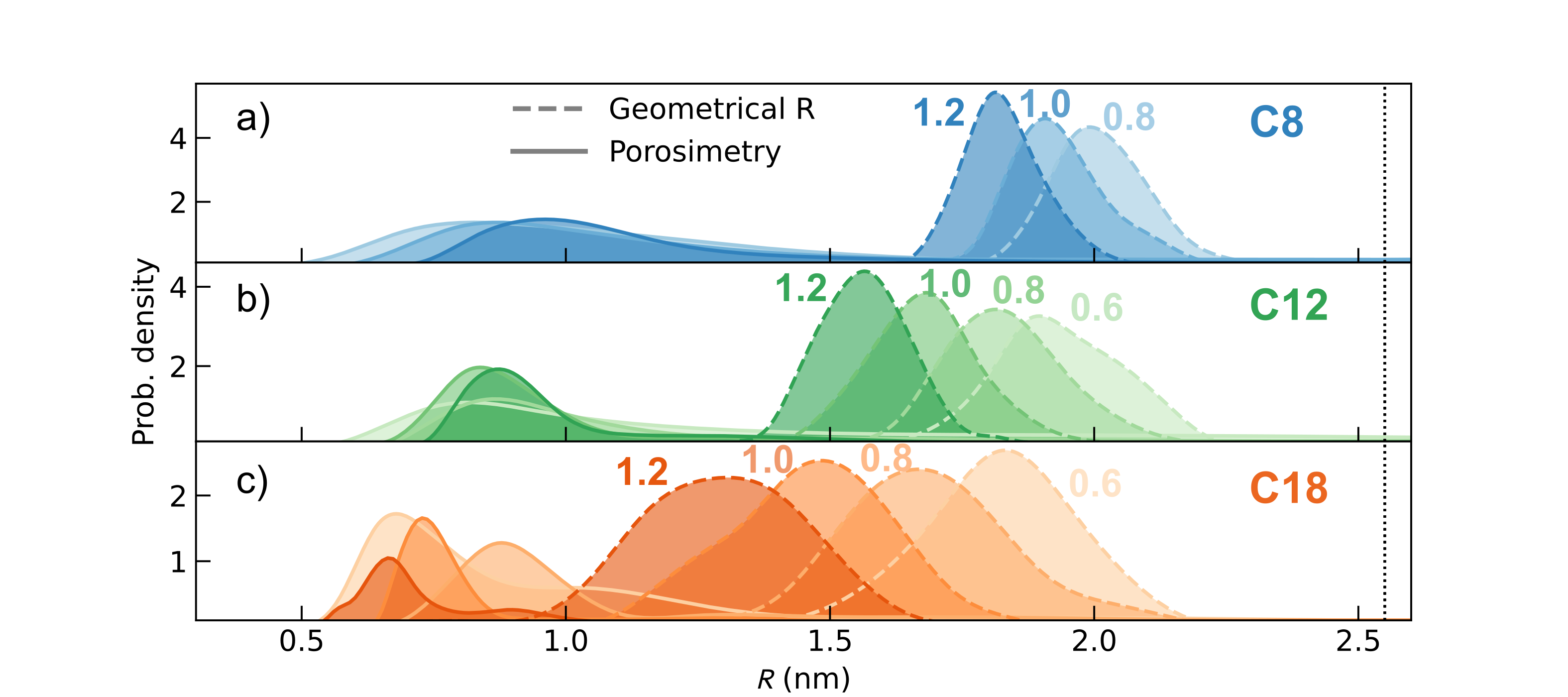}
  \caption{Comparison of the geometrical radii and radii derived from Laplace's law. Radii distributions are shown for different chain lengths, C8 (a), C12 (b), and (c) C18, as obtained from simulated porosimetry analysis based on Laplace's law (filled regions with dashed lines), compared to the actual distribution of radii as obtained by analyzing simulations (solid lines). Vertical dotted lines denote the bare pore radius. } 
  \label{Porosimetry}
\end{figure}

In porosimetry, it is generally assumed that intrusion is a deterministic process that follows Laplace's equation~\eqref{Laplace}; more precisely, it is assumed that all the pores of a given radius are filled as soon as the pressure exceeds their corresponding Laplace's pressure. Based on this assumption, the range of pressures at which intrusion takes place is interpreted as resulting from an underlying distribution of pore radii. As a result, in porosimetry, the experimental intrusion curve for a given material is used to provide information about the pore size distribution. The limitation of this procedure is that it does not take into account the variability of the radius and of the contact angle along each pore, which we just proved to be important to shape the intrusion curves. 
In our simulations, the actual distribution of local radii along the pore is a directly measurable quantity; indeed Figure~\ref{Porosimetry} show that grafted pores can have a rather broad pore size distribution with the variance, which is related to the pore roughness, increasing with grafting density.

In Figure~\ref{Porosimetry} we compare the actual pore size distribution with ``in silico porosimetry'', i.e.,  the radius distribution obtained by using Laplace's law to interpret the simulated intrusion curves in Figure~\ref{int_ext_cycles}. 
Results show that pore size distributions obtained via porosimetry vastly differ from the actual geometrical distribution.

It is particularly striking that the most probable values can differ by almost 1.5 nm for the lowest grafting densities and the variance can be either smaller or larger than the actual one depending on the chain length. In the C18 case, at 0.6 and 1.2 gps nm$^{-2}$ (Figure~\ref{Porosimetry}c), porosimetry results could imply a bimodal distribution of the radii, which is not reflected in the actual radii population. Finally, long tails corresponding to large radii apparent in the in silico porosimetry are not reflected in the actual pores, where the distribution has a well defined range. 
Overall, our analysis shows that porosimetry, in particular water intrusion porosimetry on functionalized mesopores, can significantly underestimate the mean pore size and be poorly quantitative on the overall distribution. 

In principle, also extrusion curves could be used to calculate the radius distribution through a nucleation theory; however, the stochastic nature of extrusion, together with local quantities that need to be taken into account, and the difficult independent estimates of line tension and nucleation kinetics, makes this approach more involved. 
On the other hand, Figure~\ref{Collapse_v2} shows that Laplace's equation describes quantitatively all the intrusion curves when local quantities ($R$, $\theta_Y$) are used; this result suggests that a detailed characterization of the grafting, although experimentally challenging, could significantly improve the accuracy of porosimetry.

\section*{Conclusions}
\label{sec:conc}

We performed coarse-grained MD simulations of water intrusion/extrusion in silica nanopores with different graftings. Our simulations show that grafting plays an active role in the intrusion and extrusion processes, even when the bare nanopores have an almost ideal geometry.  Depending on the chain length and the grafting density, intrusion was observed in the approximate range $3<P_\mathrm{int}<23$ MPa while extrusion was observed for $-10<P_\mathrm{ext}<12$ MPa.  

As a consequence, related quantities, such as dissipated energies, were also observed to be strongly dependent on the grafting details. A distinctive behavior was observed for intrusion and extrusion in the pore grafted with the longest hydrophobic chains at the highest grafting density considered in this work (C18 with 1.2 gps nm$^{-2}$): very pronounced slopes were observed in the intrusion/extrusion branch of the PV isotherms.

Using data collected in the simulation campaign, we were able to propose a  microscopic physical explanation of the mechanisms ruling the intrusion and extrusion processes. For intrusion, heterogeneities change the local pore radius and contact angle affecting the intrusion pressure according to a microscopically informed Laplace's law, deviating from the average Laplace's law which describes pinning at the cavity mouth only. On the other hand, extrusion is essentially a nucleation process, which is not expected to follow Laplace's law. Indeed, we find that extrusion starts by nucleation at sites within the pore where local features favor it: local constrictions were observed to play a major role in this process because they both decrease the volume of the critical bubble and they increase the local hydrophobicity. The interplay between hydrophobic and hydrophilic regions emerges as a promising aspect to be tailored for specific responses.

We discussed the specific importance of simple grafting heterogeneities on a number of relevant technologies, which include nanoporous systems for energy applications, HPLC stationary phases, and standard porosimetry protocols. In all applications of hydrophobized mesoporous materials, grafting emerges as a lead character whose fine control could provide an optimized functional design and improve characterization methods. Tailored surface graftings with complex chemistry, polymerization, and polydispersity could enable new functions, pushing towards increased efforts in the development of functionalization protocols, characterization techniques, as well as predictive theoretical tools.

\section*{Methods}
\label{sec:met}
\subsection*{Coarse-grained model}

We tuned the hydrophobicity of the matrix using different MARTINI 3 bead types. In particular, we used C1 or C2 beads in the case of the bare pores to obtain a hydrophobic surface, with water contact angles of $\theta_{Y-C1} \approx 105^\circ$ and $\theta_{Y-C2} \approx 100^\circ$, respectively. For the grafted pores, we used N1 beads for the hydrophilic silica matrix, obtaining a water contact angle of $\theta_{Y-N1} \approx 75^\circ$ 
We employed C1 and C2 beads to describe the hydrophobic carbon chain, with the C1 beads composing the ``body'' of the chain and the C2 bead the terminal methyl group. The water boxes contain equal parts of tiny, small, and regular beads, to avoid the well-known artifact of MARTINI water freezing at ambient conditions due to the presence of a surface~\cite{bruininks2019practical,zavadlav2014adaptive}.

\subsection*{Simulation protocol}
To obtain the intrusion/extrusion cycles, we performed a set of simulations increasing or decreasing the pressure by 1.5 MPa. We exploited the GROMACS COM pulling tool to apply a force to the pistons matching the required pressure. {When the pistons are pulled against each other, the imposed pressure is positive; when they are pulled apart, the imposed pressure is negative.}
We ran the simulations in the NVT ensemble using a (stochastic) V-rescale thermostat. During the cycle, the beads of the matrix were frozen in all three dimensions, while the pistons were free to move only along the cylinder axis and frozen in the other two directions. We used a Verlet cutoff to update the neighbor list in combination with a straight cutoff of 1.1 nm and a potential shift to zero at the cutoff distance. We ran the simulations with a time step of 20 fs for 10 ns at each different pressure.
Five independent simulations were performed for each of the systems with different features, starting each run by changing the grafting of the initial configuration. 
To compute the contact angles of water on grafted silica surfaces, we performed a set of 5 NVT simulations (each 1 $\mu$s long), starting from different initial grafting configurations. We used a leap-frog stochastic dynamics integrator with the same time step as before. 
More details on the simulation protocols and calculation methods can be found in the Supplementary section "Supplementary Methods" (Supplementary Figures~S1,S2,S3,S4,S5,S7).

\section*{Acknowledgements}

Funded by the European Union (ERC-PoC, NODRY, 101059685). Views and opinions expressed are however those of the authors only and do not necessarily reflect those of the European Union or the European Research Council Executive Agency. Neither the European Union nor the granting authority can be held responsible for them. The project leading to this application has received funding from the European Union’s Horizon 2020 research and innovation program under grant agreement No 101017858.
This article is part of the grant RYC2021-032445-I funded by MICIN/ AEI/10.13039/501100011033 and by the European Union NextGenerationEU/PRTR.

\section*{Competing interests}

The authors declare no competing interests. 

\section*{Contributions}

R.F, C.S., T.A., B.D., G.A. and R.G. wrote the paper. R.F., C.S. and B.D. analysed the data. G.A., R.G. and G.Y. conceived the core idea of the paper and framed the research. G.Y. provided experimental consultancy. G.A. T.A., and G.R. supervised research. All authors critically revised the manuscript.

\section*{Data Availability}

The data shown in the figures, the scripts used for analysis and a selection of full molecular dynamics trajectories are available in Zenodo with the identifier 10.5281/zenodo.10379810. Any additional data can be provided upon request.

\bibliography{bib}

\end{document}


\maketitle

\begin{table}[]
\footnotesize

\begin{adjustbox}{width=\textwidth}
\begin{tabular}{ccccccccc}

\hline
\multicolumn{2}{l}{System} & $\langle R \rangle$ (nm) & $R_\mathrm{min}$ (nm) & $\theta_Y$ ($^\circ$) & $\theta_{Y, \mathrm{max}}$ ($^\circ$) & $P_\mathrm{int}$ (MPa) & $P_\mathrm{ext}$ (MPa) & $E_d$ (J/g) \\ \hline
\multirow{3}{*}{C8}  & 0.8 & 1.977$\pm$0.003     & 1.89$\pm$0.02     & 95$\pm$3              & 98                     & 5$\pm$1         & -12.1$\pm$0.8               &      2.4$\pm$0.1      \\
                     & 1.0 & 1.928$\pm$0.003     & 1.82$\pm$0.05     & 98$\pm$3              & 102                    & 6.8$\pm$0.4     & -9.9$\pm$0.7    &   2.64$\pm$0.08     \\
                     & 1.2 & 1.857$\pm$0.001     & 1.73$\pm$0.02     & 102.1$\pm$2.5         & 109                    & 9$\pm$1         & -5.9$\pm$0.7    &    2.3$\pm$0.2    \\ \hline
\multirow{4}{*}{C12} & 0.6 & 1.922$\pm$0.003     & 1.75$\pm$0.07     & 97.8$\pm$2.5          & 101                    & 6.8$\pm$0.6     & -8.6$\pm$1.     &      2.49$\pm$0.04     \\
                     & 0.8 & 1.861$\pm$0.005     & 1.71$\pm$0.08     & 102$\pm$2             & 105                    & 10.3$\pm$0.6    & -5.9$\pm$0.7    &   2.3$\pm$0.2     \\
                     & 1.0 & 1.752$\pm$0.002      & 1.56$\pm$0.09     & 105$\pm$1             & 113                    & 12.7$\pm$0.8    & -1.1$\pm$0.6    &      1.9$\pm$0.1       \\
                     & 1.2 & 1.663$\pm$0.005     & 1.50$\pm$0.03     & 108$\pm$2             & 111                    & 16$\pm$1        & 2.4$\pm$0.4     &      1.9$\pm$0.1       \\ \hline
\multirow{4}{*}{C18} & 0.6 & 1.829$\pm$0.003     & 1.63$\pm$0.06     & 100.6$\pm$1.5         & 106                   & 10.5$\pm$1.4    & -3.9$\pm$0.8    &       2.3$\pm$0.2      \\
                     & 0.8 & 1.733$\pm$0.002     & 1.45$\pm$0.05     & 108$\pm$3             & 111                    & 15$\pm$1        & 0.6$\pm$1.2     &     2.1$\pm$0.2        \\
                     & 1.0 & 1.572$\pm$0.007     & 1.24$\pm$0.06     & 111$\pm$4             & 113                & 20$\pm$1        & 6$\pm$1         &       1.8$\pm$0.2      \\
                     & 1.2 & 1.420$\pm$0.005     & 1.10$\pm$0.05     & 112.6$\pm$4.5         & 114                & 24$\pm$3        & 14$\pm$2        &      1.9$\pm$0.4       \\ \hline

\end{tabular}

\end{adjustbox}

\caption{Summary of the main quantities characterizing the grafted pores considered in this simulation work, including the mean radius $\langle R \rangle$, the minimal one $R_\mathrm{min}$, the mean contact angle $\theta_Y$, and the maximum one $\theta_{Y,\mathrm{max}}$, the intrusion pressure $P_\mathrm{int}$, the extrusion one $P_\mathrm{ext}$, and the dissipated energy $E_d$. {We remind that pressures and energies should be multiplied by 3.41 to translate these quantities to experimental values.}}
\label{table_energy}

\end{table}

\section*{Supplementary Methods}

\subsection*{Functionalization of the CG nanopore}
    \begin{figure}[H]
        \centering
        \includegraphics[width=0.7\linewidth]{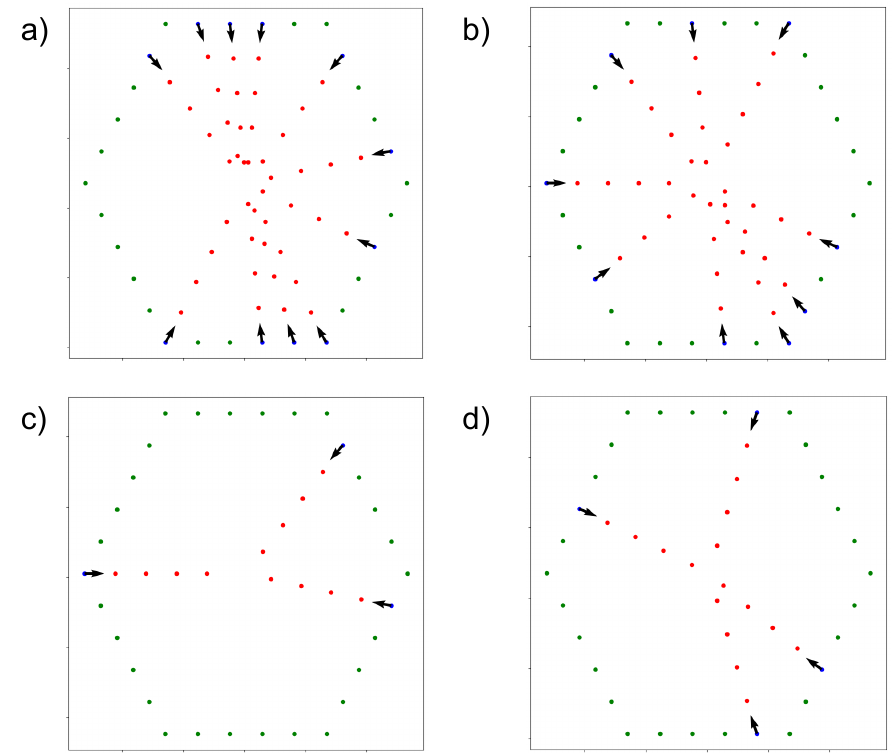}
        \caption{Cross-section of the nanopores at different z coordinates along its axis.The green points are the beads belonging to the layer of the solid matrix in contact with water. In blue, we represented the beads randomly selected to be the grafting sites. The ligands (in red) are bonded to the surface beads. Panels a) and b) are two sections of a pore grafted with C18 chains with grafting density of 1.2 grps/nm$^2$, while the sections in c) and d) refer to a grafting density of 0.6 grps/nm$^2$ with C18 chains.} 
         \label{SI-silanization}
    \end{figure}

\subsection*{Contact angle and SASA}
    The simulated systems consist of a $500 \ \textrm{nm}^2$ fcc surface made of N1 beads and different grafting, and a $4000 \ \textrm{nm}^3$ cylindrical water droplet, composed of the three different Martini sizes (R, S, T) in equal number percentage.
    We performed 1 $\mu$s CG simulations in the NVT ensemble at 298 K to compute the contact angle of water on silica surfaces with different functionalizations. For each system, we performed five runs with different starting configurations. In every run, the contact angle was calculated
    from the last 500 ns of the trajectory.
    To compute the contact angle, we used an in-house Python script that performs a circular cap fit to the isodensity contour of the border of the droplet with the vacuum interface. The same simulation and analysis protocols were used to compute the contact angle on the bare hydrophobic surfaces, made of C2 and C4 beads.
  
    The Solvent Accessible Surface Area (SASA) was calculated using the GROMACS tool sasa. We used the smallest bead radius in our water model (0.191 nm) as the radius of the probe. For each grafting density and chain length, the results are averaged over all the realizations of the pore.
    
\subsection*{Intrusion/extrusion cycles}

\begin{figure}[H]
    \centering
      \includegraphics[width=1\linewidth]{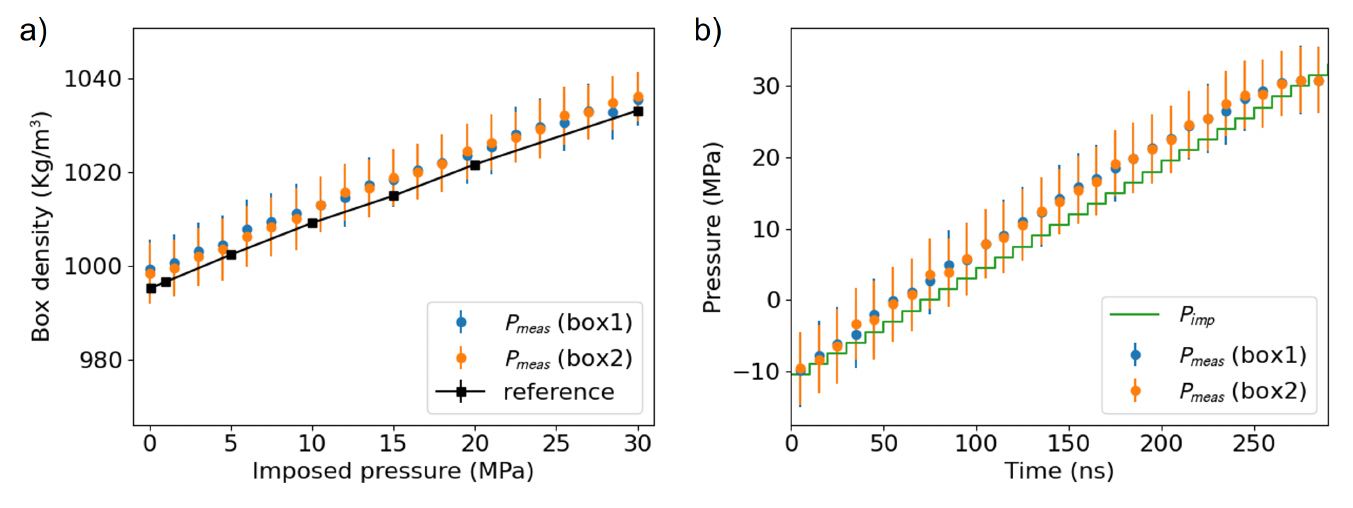}
      \caption{{a) Comparison between water densities in the two reservoirs (reservoirs 1 and 2) at different pressures during the intrusion and the density in the NPT water box (reference). b) Comparison betweeen the nominal pressure (in green) and the pressures derived from water densities in the two reservoirs.}} 
      \label{SI-water_dens}
    \end{figure}

\begin{figure}[H]
    \centering
      \includegraphics[width=1\linewidth]{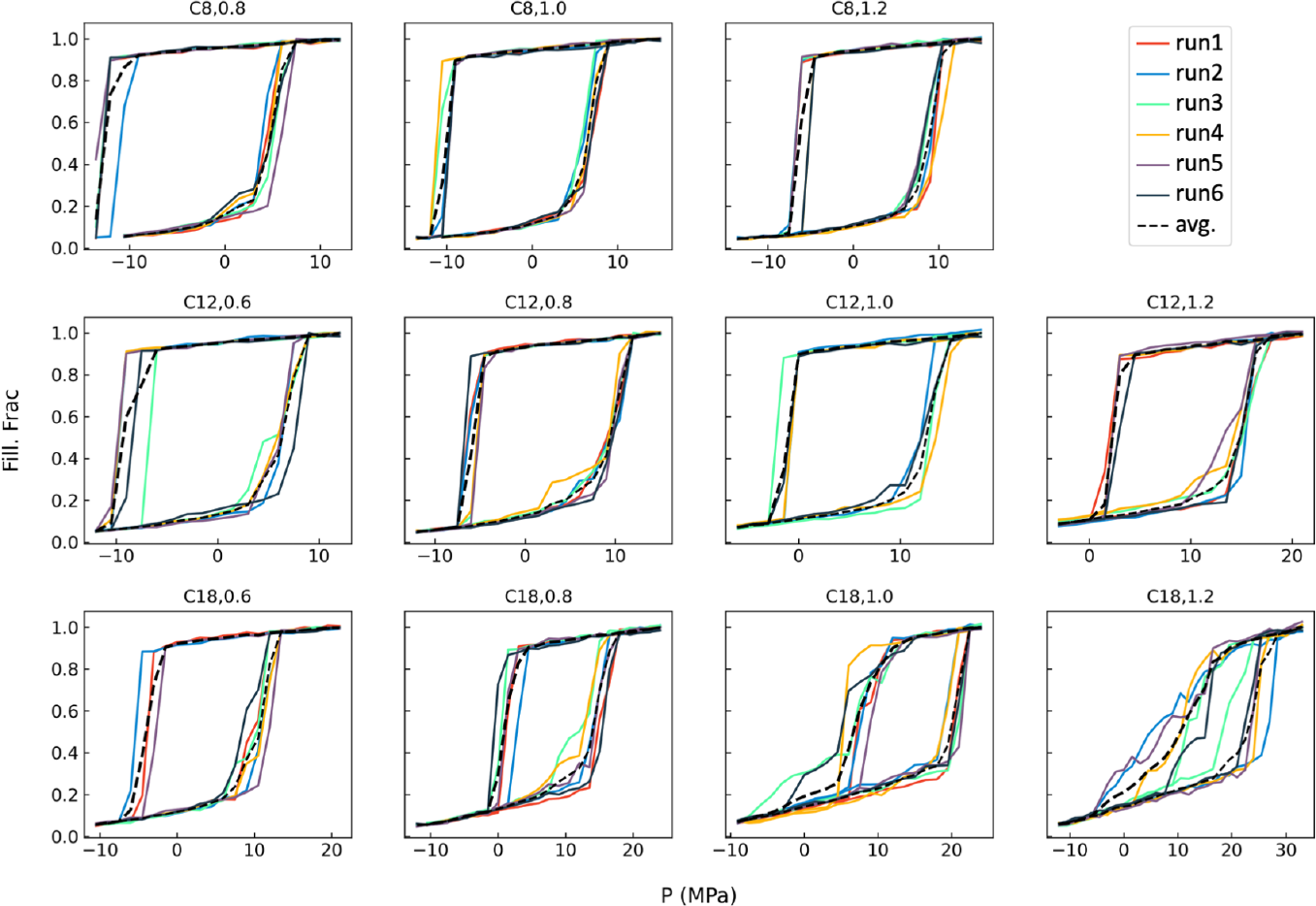}
      \caption{Intrusion/extrusion for all single realizations of the pores and their respective average.} 
      \label{SI-all_runs}
    \end{figure}

    To compute intrusion/extrusion cycles, two movable pistons were used. The pistons consist of a thin fcc lattice with dimensions (20x20x2.5) $\textup{nm}^3$. The length of the nanopore is 20nm, and the radius is 5.2nm previous to any grafting. The starting dimensions of the two water cubic boxes are (20x20x20)$\textup{nm}^3$. We used an in-house Python script, exploiting the MDAnalysis library\cite{michaud2011mdanalysis,gowers2016mdanalysis}, to analyze our simulations.
    Firstly, we computed the density of the water reservoirs during the I/E cycle to monitor the pressure applied by the pistons on the system. We calculated the densities of two water boxes located at a fixed distance from the pistons at each pressure step. Then we compared the results with those obtained for a water box with the same bead size composition, simulated in the NPT ensemble using a Parrinello-Rahman barostat. As shown in Fig.~\ref{SI-water_dens}, the three sets of values match, proving the proper functioning of our pulling protocol.
    The same script was used to compute the filling of the pore, calculating the average number of intruded water beads at each pressure for every run. We then derived the filling fraction by dividing the number of intruded beads by the maximum filling of each run.\\
    The I/E cycles for all the single realizations of the pores are reported in Fig. \ref{SI-all_runs}.

The intrusion and extrusion pressures were defined as the inflection point of the intrusion/extrusion curve. This definition is used for all the analysis present in the text, except for the confront with nucleation theory (Fig. 7), where extrusion was defined to be the point where the emptying of the pore starts, more precisely, the point where the filling fraction drops below the average filling of the last 5 points.

\subsection*{Mean thickness, $\langle R \rangle$ and  $\theta_{Y,max}$}
\begin{figure}[H]
    \centering
      \includegraphics[width=0.8\linewidth]{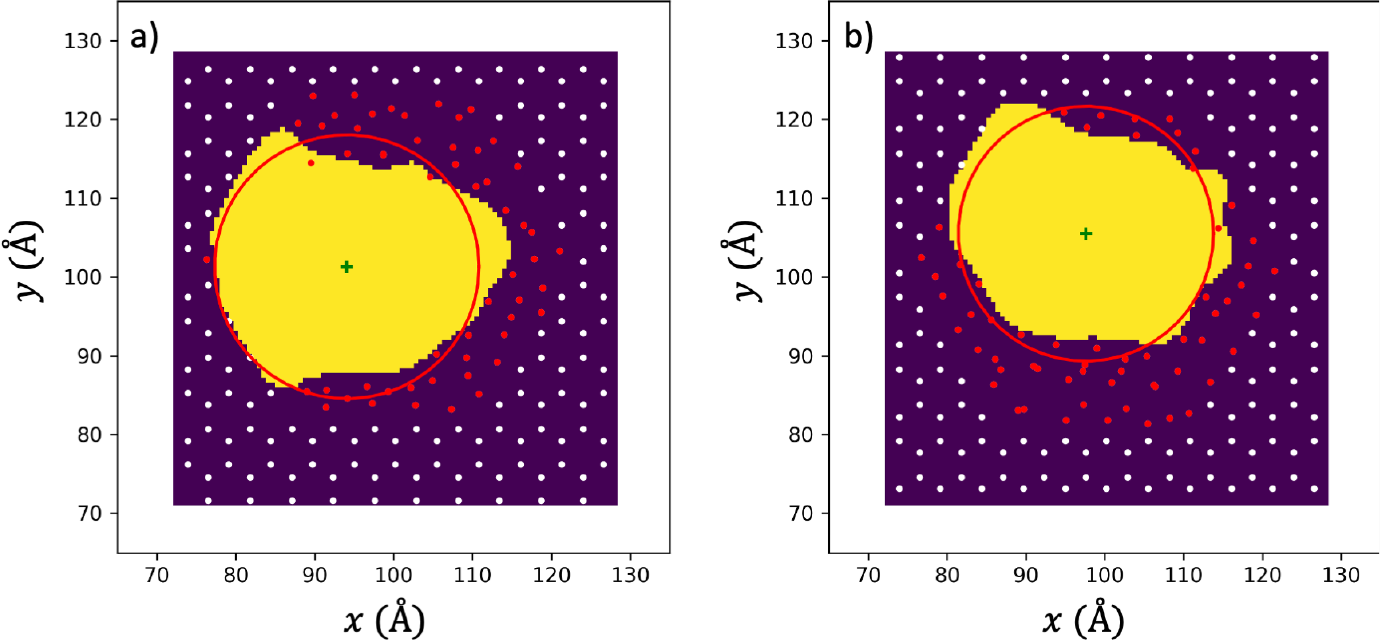}
      \caption{a) and b) represent two different sections of the same pore (a realization of C18,0.6). The red and white points are beads from the silane molecules and the substrate, respectively. The mean radius is derived from the yellow area.} 
      \label{SI-eff_radius}
    \end{figure}

\begin{figure}[H]
    \centering
      \includegraphics[width=0.5\linewidth]{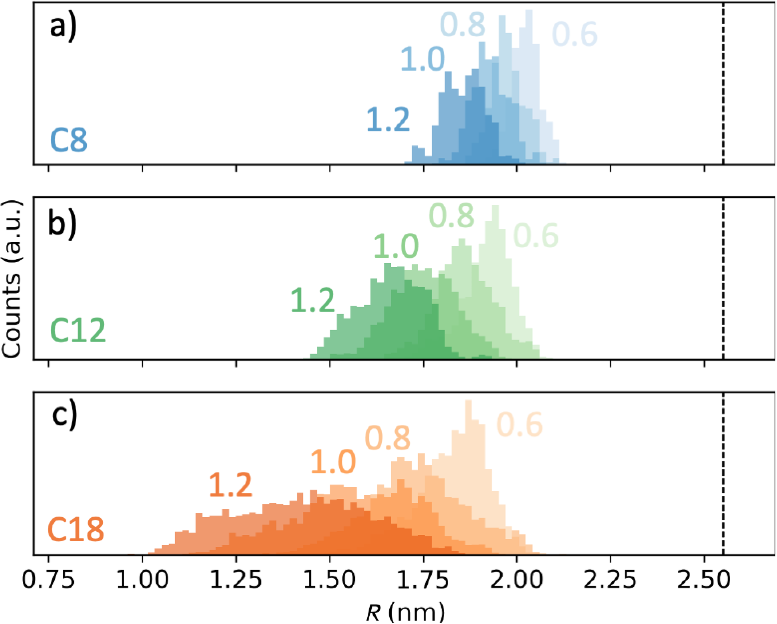}
      \caption{Radius distributions for a) C8, b) C12 and c) C18. The dashed line represents the bare pore radius.} 
      \label{SI-radius_dist}
    \end{figure}

\begin{figure}[H]
    \centering
      \includegraphics[width=1\linewidth]{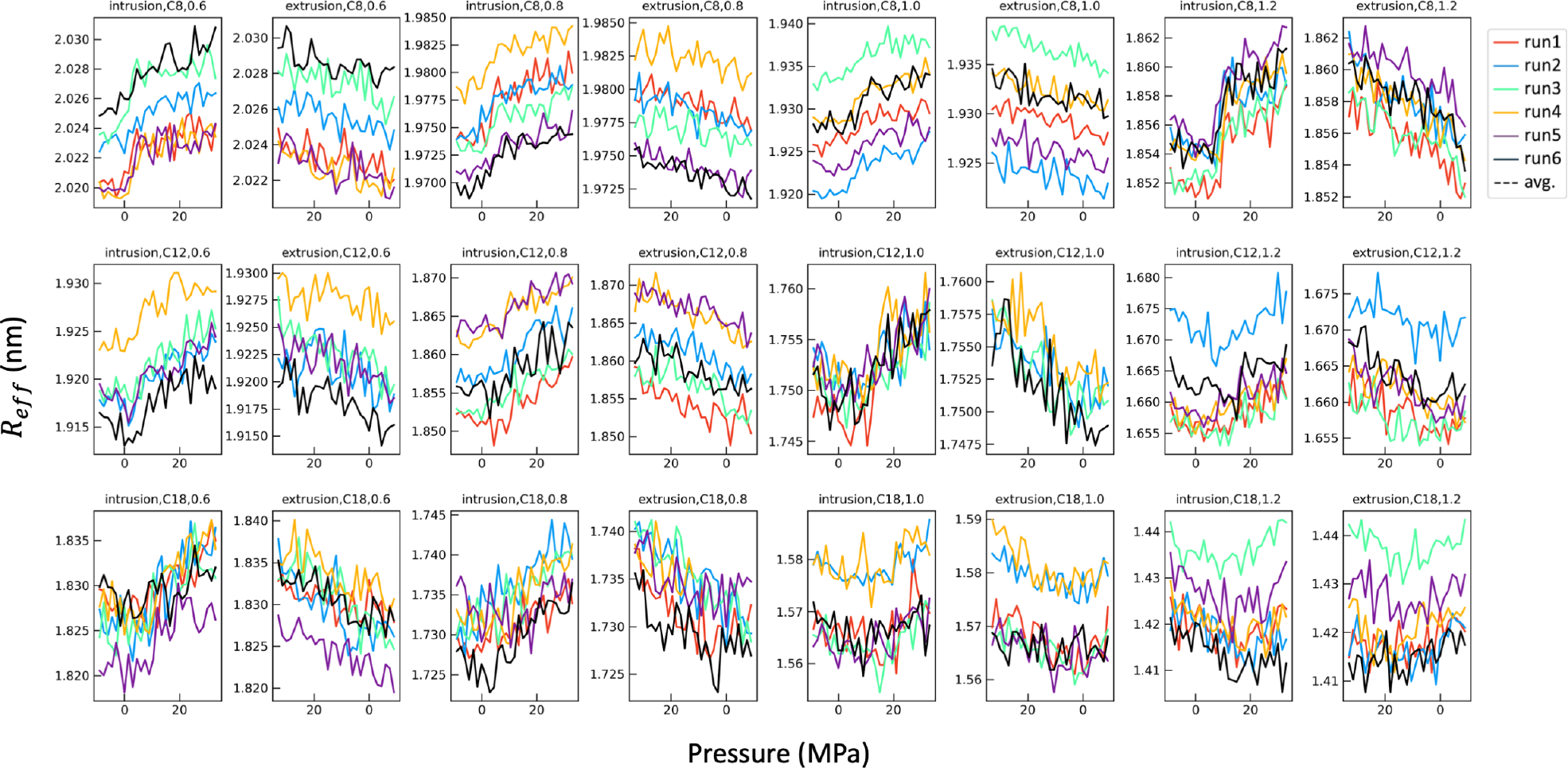}
      \caption{mean radius in function of pressure for each system during intrusion and extrusion. The radius of all systems increases/decreases around 0.01 nm during intrusion and extrusion.} 
      \label{SI-compressibility}
    \end{figure}

    The mean radius was calculated by dividing the pore into 20 cross-sections and performing a 2D kernel density estimation using the beads present between two consecutive cross-sections. Then, by selecting only the region below a low density threshold, we obtain the empty area of the pore section, from which the mean radius is derived (Fig.~\ref{SI-eff_radius}). The mean thickness for each section is calculated by simply subtracting the mean radius from the approximate radius of the bare pore (2.55 nm). Fig.~\ref{SI-radius_dist} shows the radius distributions for each chain length and grafting density. Fig.~\ref{SI-compressibility} shows the variation of the mean radius with the pressure.

    \begin{figure}[H]
    \centering
      \includegraphics[width=0.8\linewidth]{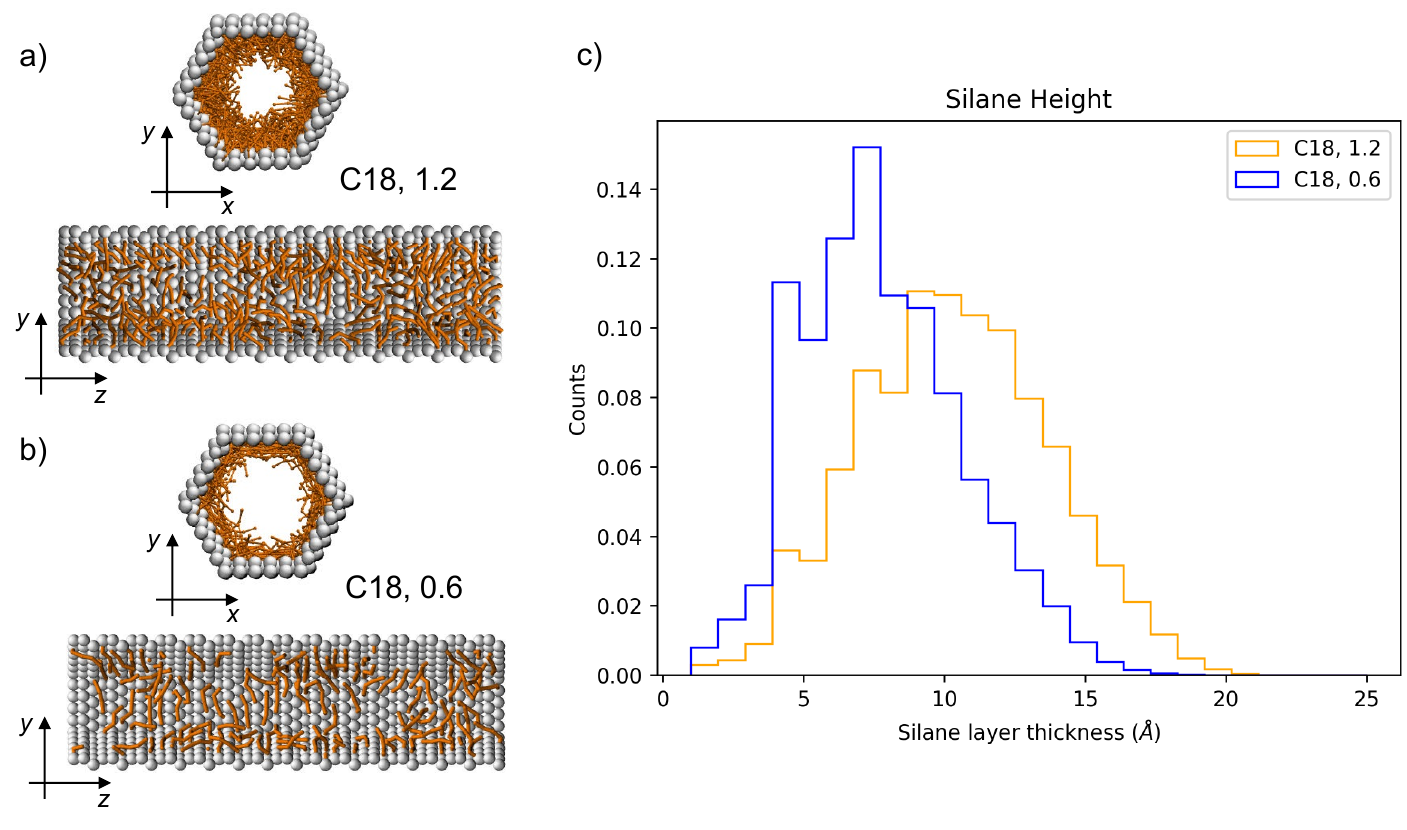}
      \caption{Arrangement of the grafted molecules (C18) at different densities (0.6 and 1.2 grps/nm$^2$). Panel a) and b) show x-y (upper insets) and y-z (lower insets) sections of the nanopores. Panel c) reports the distributions of thickness of the grafting molecules in the two cases, showing that at lower density the hydrophobic ligands tend to lie on the surface, resulting in a thinner layer of grafted molecules, while at higher density, the molecules are more compact, resulting in a thicker ligand layer. Given a fixed chain length, the arrangement of the ligands changes, resulting in layers of different thickness. However, the arrangement of chains appears to always remain disordered.} 
      \label{SI-grafting-arrangement}
    \end{figure} 

    \begin{figure}[H]
    \centering
      \includegraphics[width=0.6\linewidth]{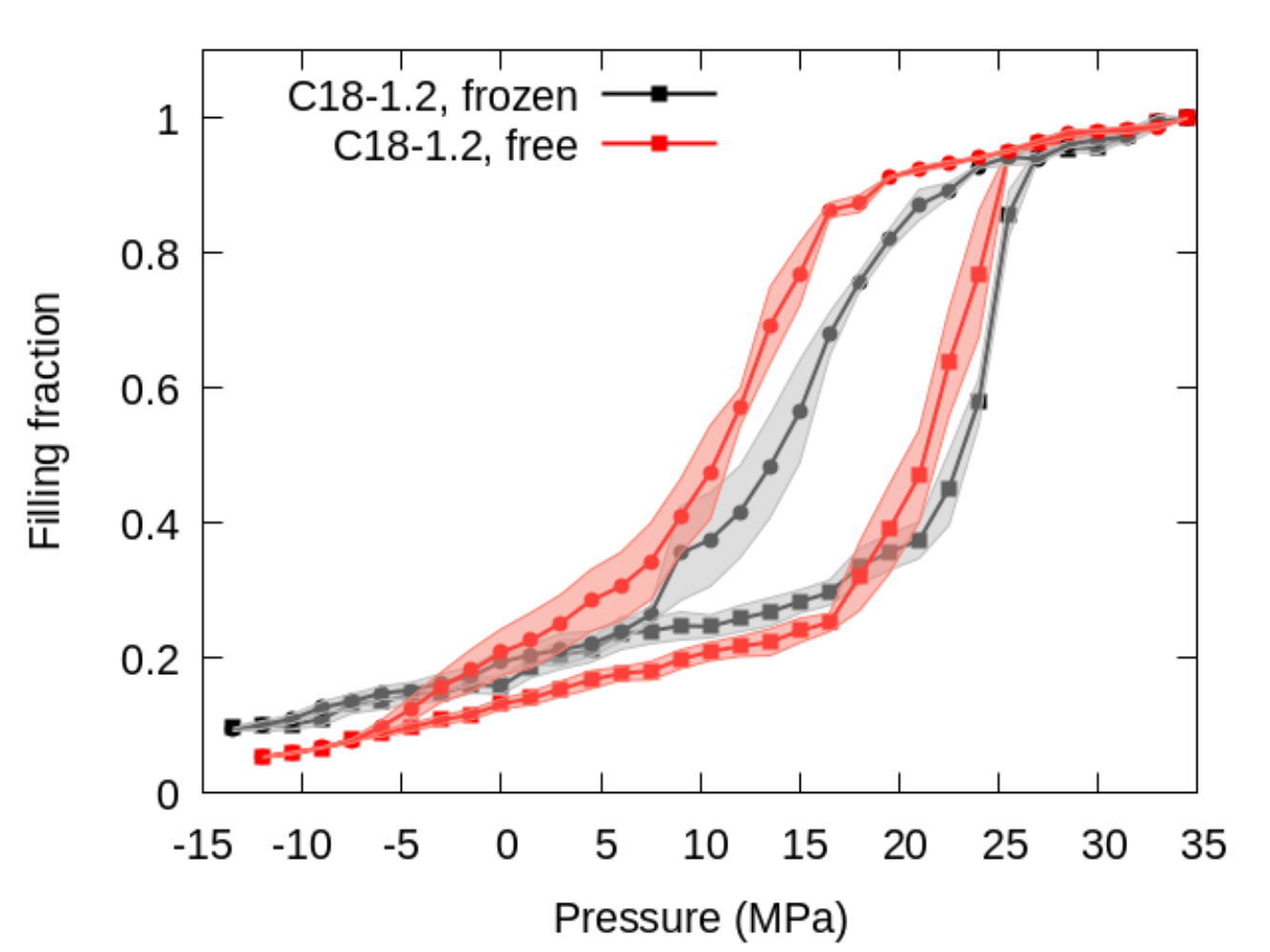}
      \caption{{I/E cycles for C18-1.2 systems. The red curve refers to the systems with mobile ligands and the black one to the systems with immobilized ligands. The intrusion ramps are almost coincident, while the discrepancy in the extrusion ramps can be considered negligible, given the errors.}} 
      \label{SI-freeze-sil}
    \end{figure} 

    The maximum contact angle estimates the presence and size of hydrophobic patches on the pore surface. We divided the channel into 8 bins along the cylinder axis direction, and, in each slice, we computed the ratio between the hydrophobic and hydrophilic beads composing the surface of the pore. Then, we calculated the grafting density corresponding to each ratio. In this way, it was possible to identify the most hydrophobic slice and the corresponding contact angle. We averaged the maximum contact angle values obtained for the five different initial configurations of each system.

\subsection*{Dissipated energy}
    The dissipated energy is calculated by integrating the area between intrusion and extrusion curves and multiplying it by the pore volume per gram of a typical MCM-41 matrix. Here we used the value of 330 mm$^3$/g from Guillemot et al.\cite{guillemot2012}
    
\subsection*{Surface tension}
    The surface tension was calculated by using the test area method \cite{vega2007surface}. The method consists of estimating the change in free energy associated with a small change in the interfacial area at constant volume. The obtained value is $\gamma_\mathrm{lv}=21\pm1$~mN/m.

\subsection*{Porosimetry}
    A common practice to assess the population of pores via water/mercury porosimetry is to take the derivative of the PV isotherm measured during the intrusion, which gives a distribution peaked at the intrusion pressure, and map it into a radius distribution through the Laplace equation. We carried out this procedure by using the Young contact angle as obtained from simulations of a droplet over a flat substrate with the same grafting density and chain length as the respective pore. It is worth remarking that, in practice, the real grafting density should be affected by the curvature of the pore.

\section*{Bare pore cycles}
\begin{figure}[ht!]
    \centering
      \includegraphics[width=0.6\linewidth]{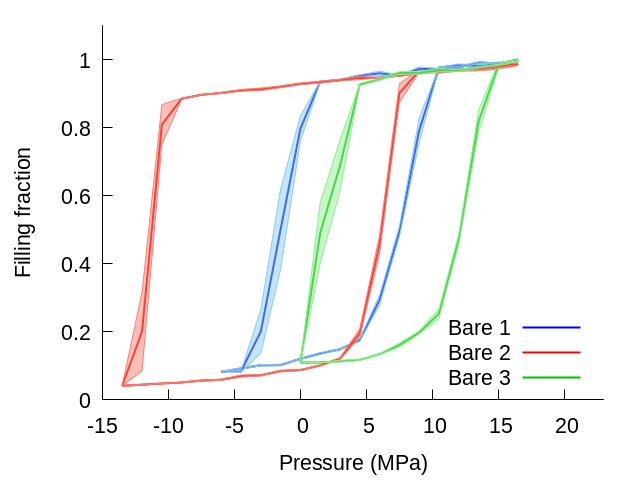}
      \caption{IE cycles of three bare pores with different hydrophobicities and radii (Bare 1: $\theta_{Y}=100^\circ$ and $R=1.3 \ \textrm{nm}$, Bare 2: $\theta_{Y}=100^\circ$ and $R=1.8 \ \textrm{nm}$, and Bare 3: $\theta_{Y}=105^\circ$ and $R=1.3 \ \textrm{nm}$). As expected, given the same hydrophobicity of the surface (Bare 1 and Bare 2), both the intrusion and extrusion curves shift to lower pressures as the radius increases (Bare 2). On the other hand, if the radius is the same (Bare 1 and 3), the system with higher hydrophobicity (Bare 3) has higher I/E pressures.}
      \label{SI-bare_pores}
    \end{figure}

    As benchmark tests, we modeled three bare channels with different radii and hydrophobicities. Fig.~\ref{SI-bare_pores} shows the results of the I/E cycles.
    
\section*{Single intrusion events}

    \begin{figure}[H]
    \centering
      \includegraphics[width=0.6\linewidth]{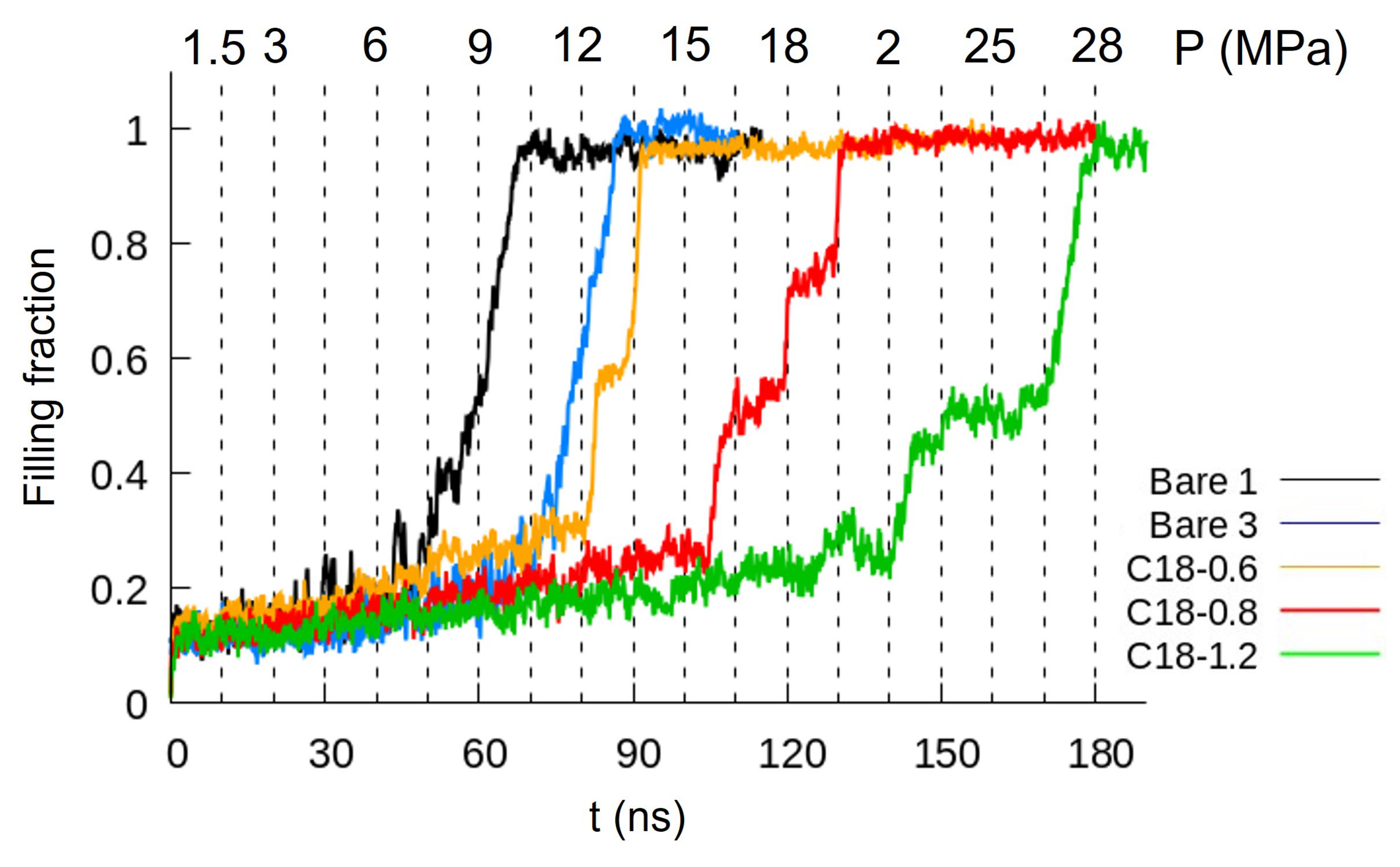}
      \caption{Comparison between intrusion ramps of bare channels (Bare 1 and Bare 3) and C18-grafted pores (C18-0.6, C18-0.8, C18-1.2). The curves are relative to single realizations of the systems. Increasing the grafting, the slope becomes less sharp, and the number of steps, distinctive of the stick-slip behavior due to the presence of the ligands, increases.} 
      \label{SI-slopes}
    \end{figure}

{\section*{Free-energy calculations}}

    \begin{figure}[ht!]
    \centering
      \includegraphics[width=1\linewidth]{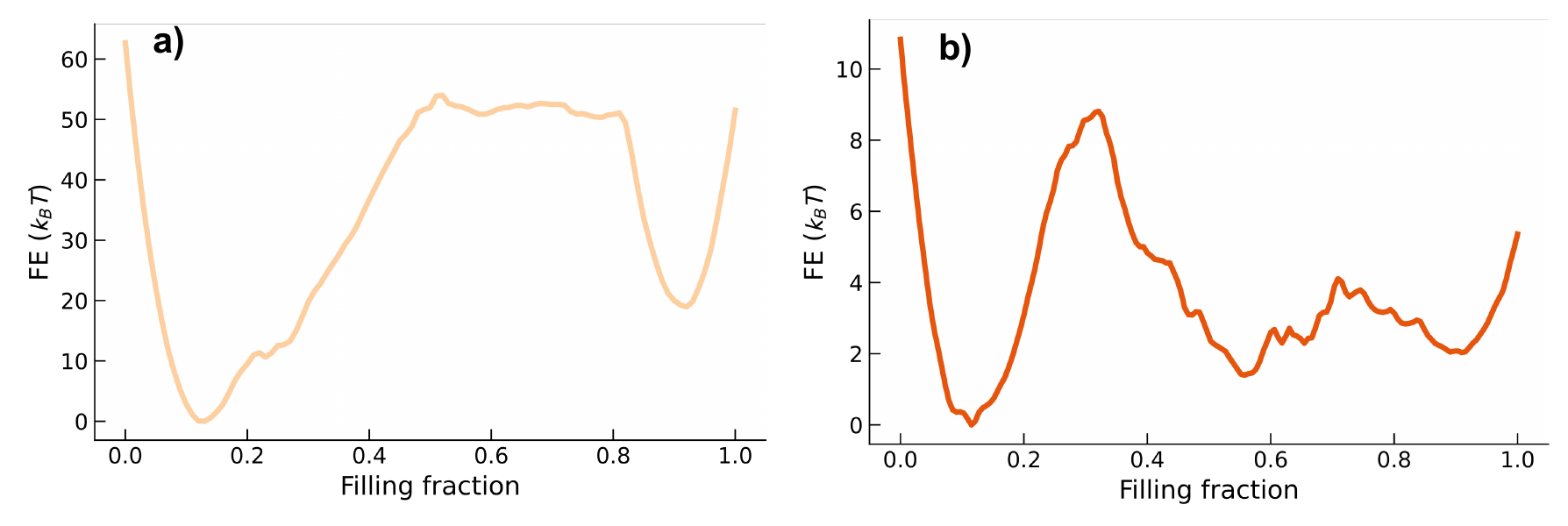}
      \caption{{Free energy profiles close to intrusion pressure for C18 with grafting densities of a) 0.6 gps/nm$^2$ and b) 1.2 gps/nm$^2$. The lowest grafting density case shows a deeper minimum for the filled state, given that it has more hydrophilic area exposed. It is interesting to notice the presence of intermediate minima due to the heterogeneities in the grafting layer; this effect is especially clear for the highest grafting density.}} 
      \label{free_energy}   
    \end{figure}
    
{We performed the calculation of the free energy profile for water intrusion, using Restrained Molecular Dynamics (RMD). RMD is based on the idea of performing simulations in which a collective variable is biased using a harmonic potential to yield an estimator for the derivative of the free energy with respect to the collective variable at a particular value of the variable itself. The method is not dissimilar from to umbrella sampling and was proposed in Ref~\cite{TAMD} (Eq.~12). Filling of the pore was used as the relevant collective variable to describe the transition of the pore from the wet to the dry phase. A differentiable estimator for the pore filling was obtained by identifying the pore filling with the (smoothed) Coordination Number of the water beads calculated with respect to the geometrical center of the pore. More specifically, our model for MARTINI water is composed of three different bead sizes, namely regular (R), small (S), and tiny (T). A linear combination of the three coordination numbers for each bead size was used as the collective variable describing the volume filling fraction of the pore. The coefficients which were used in the linear combination of the coordination numbers for the R, S, and T beads are: $c_{R}=1.43788,c_{S}=1.18632,$ and $c_{T}=0.50987$. These values were obtained by analyzing the bead populations during long unrestrained filled pore simulations. This procedure relies on the assumption that, for each sampled MD frame, the same geometrical volume was attained by the different bead populations that were observed inside the pore. Biasing was obtained, in all free energy simulations, by using the PLUMED 2 software \cite{plumed} as a plugin for the GROMACS MD package. During simulations, the working pressure was set at 1.5 MPa. Free energies at different pressures can then be estimated with good approximation by tilting the free energy profile, subtracting a $\Delta PV_{v}$ that is linear in the vapor volume. The resulting profiles are shown in Fig.~\ref{free_energy}, and they correspond to the free energy profiles at the intrusion pressure. Coherently with the microscopic insights offered by unbiased simulations, the profiles are characterized by intermediate free energy barriers that are caused by the heterogeneities in the grafting density along the pore.}

\section*{Nucleation sites}
    {Videos of nucleation processes, as well as for intrusion trajectories can be found at the following link: \href{https://zenodo.org/uploads/10379810}{Trajectories repository}}
    
    \begin{figure}[H]
    \centering
      \includegraphics[width=0.9\linewidth]{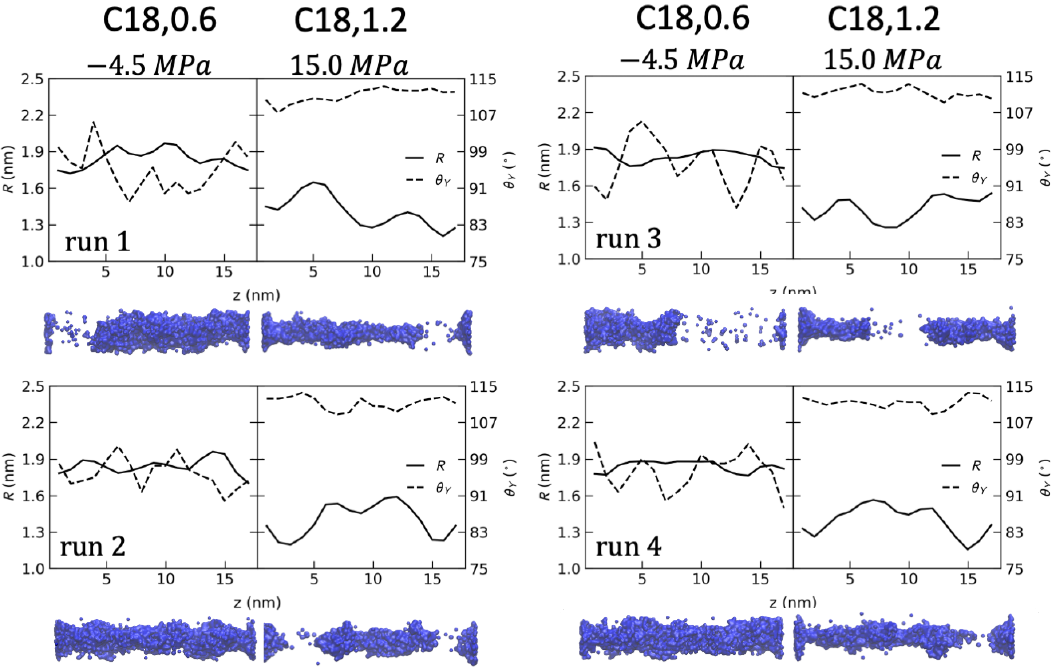}
      \caption{Local radius of the grafted pore and local contact angle $\theta_Y$ as a function of the axial coordinate for two representative systems with low (C18-0.6) and high (C18-1.2) grafting densities and 4 different realizations of the pore. The MD snapshots show the water inside the pore when the pressure is nearby the extrusion pressure. While, at low grafting densities, the nucleation site is not always evident, at high grafting densities the coordinates where the vapor phase nucleation occurs match the ones of the minimum radius for every single run.} 
      \label{SI-nucleation}
    \end{figure}

    \begin{figure}[H]
        \centering
        \includegraphics[width=0.9\linewidth]{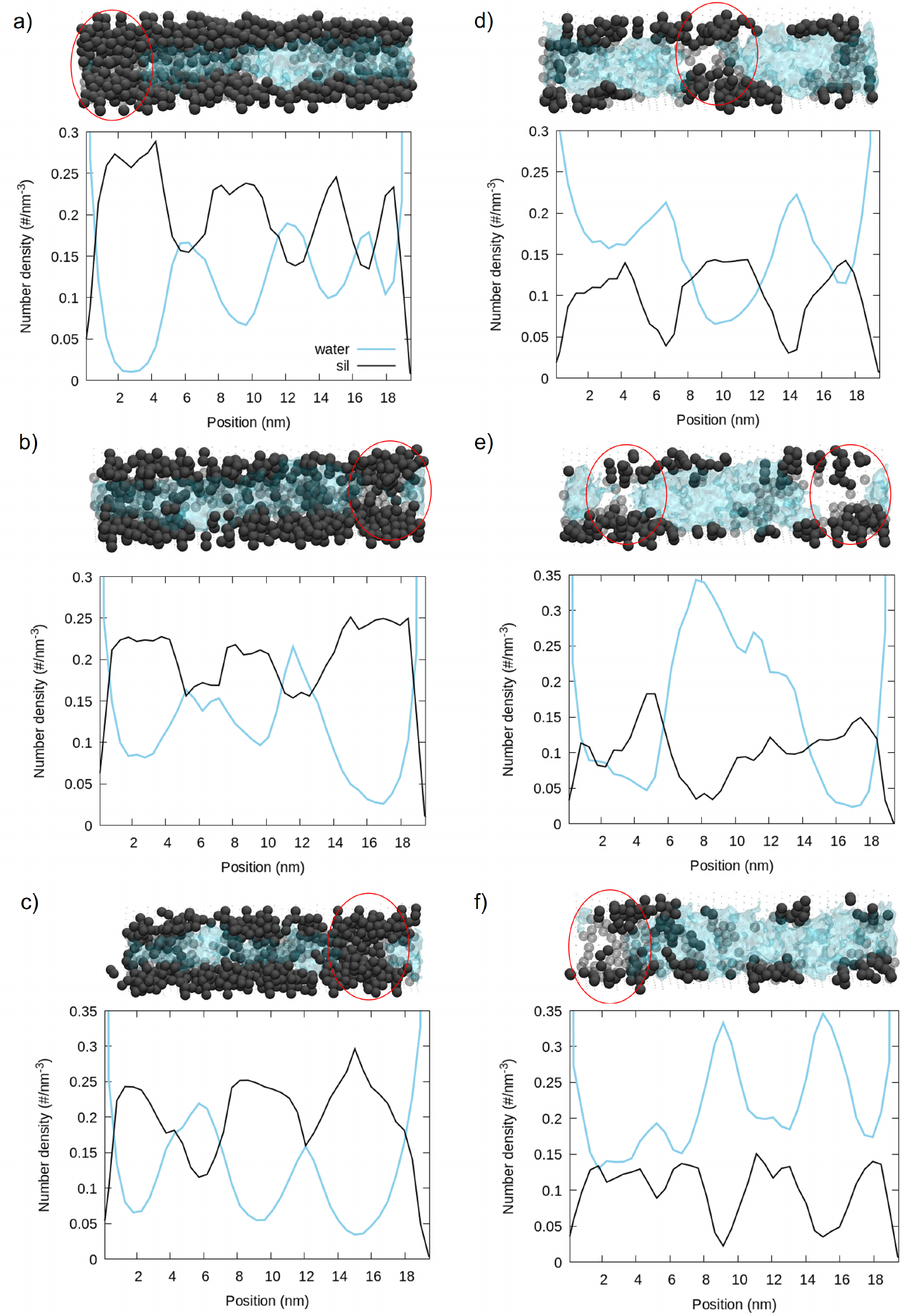}
        \caption{{Correlation between nucleation events (in the insets) and water and silane number density along the pore for two representative systems at high (C18-1.2, in panels a), b), and c)) grafting densities with low (C18-0.6, in panels d), e), and f)) and 3 different realizations of the pore. The nucleation sites correspond to maxima of the silane density and minima of the water density.}}
        \label{nucl_vs_dens}
    \end{figure}

    {\begin{figure}[H]
        \centering
        \includegraphics[width=\linewidth]{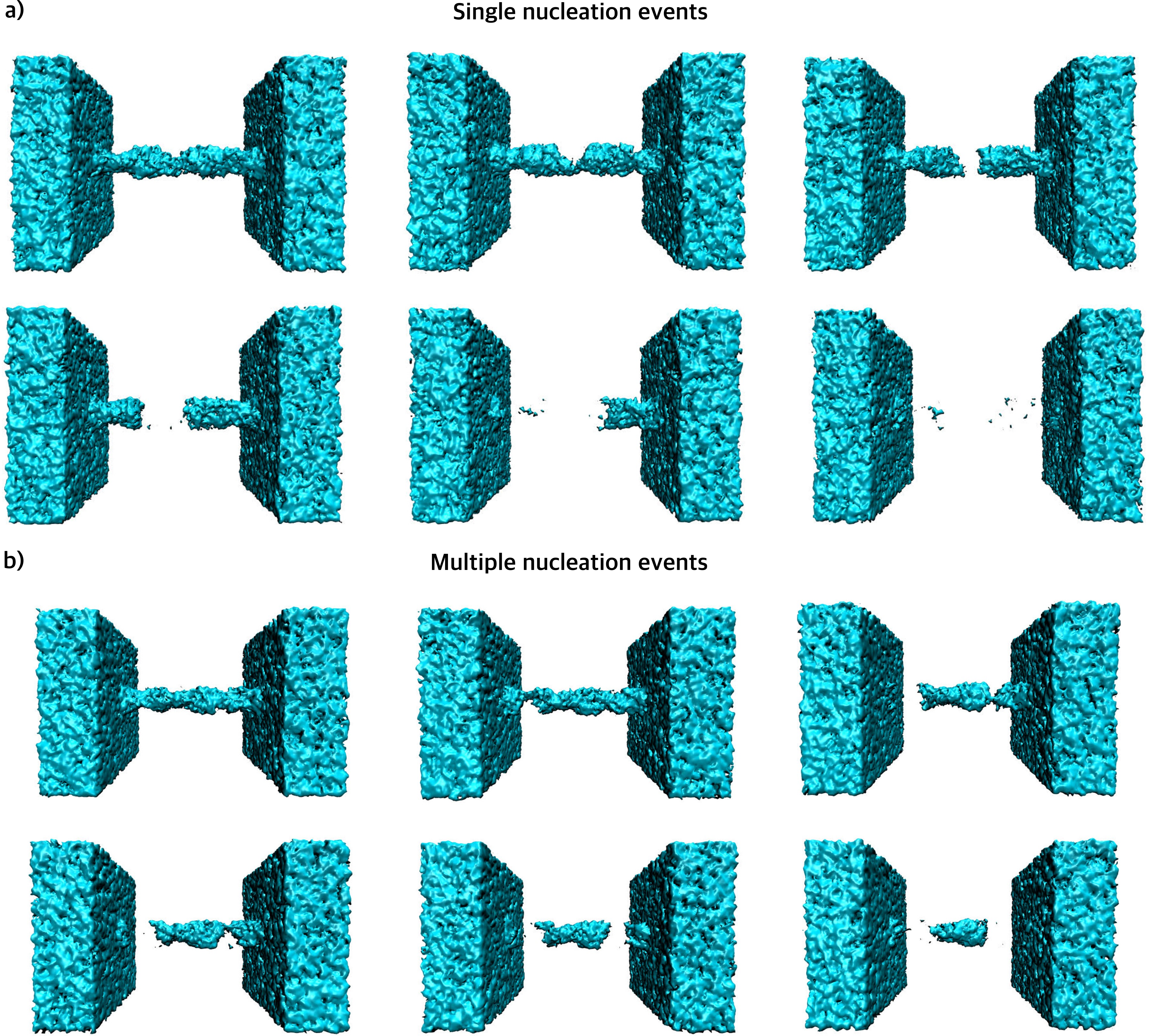}
        \caption{{Snapshots of the formation of a) single or b) multiple nucleation sites and consequent emptying of the pore in two different C18-0.6 realizations.}}
        \label{nucl_sites}
    \end{figure}

    \begin{figure}[ht!]
    \centering
      \includegraphics[width=.6\linewidth]{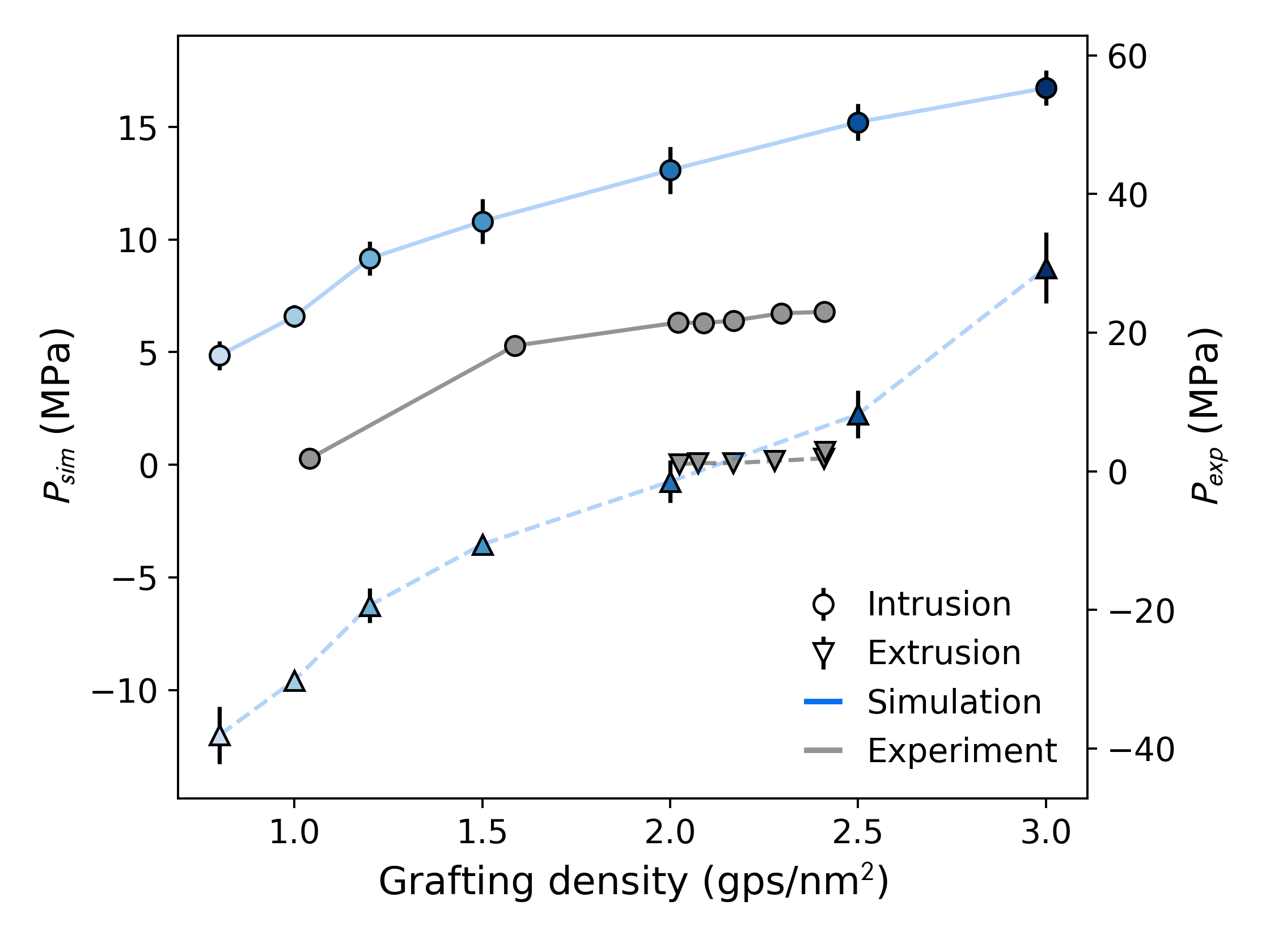}
          \caption{Intrusion (circular points) and extrusion (triangular points) pressures obtained from simulation (blue) and from experiments~\cite{fadeev1997study} (gray). $P_\mathrm{sim}$ is the pressure directly obtained from simulations, while $P_\mathrm{exp}$ is the pressure after applying a rescaling to correct for the surface tension difference (pressures were multiplied by $\gamma_\mathrm{exp}/\gamma_\mathrm{sim}$).} 
      \label{SI-Pint_Pext}
    \end{figure}

{\section*{Bridging simulations and experiments: effect of time and length scales on experimental and simulated I/E pressures}}
{\subsection*{Extrusion pressure}}
{
Extrusion is ruled by the nucleation of a vapor bubble within the pore. The barrier for the nucleation of such a bubble is given by eq.~(2) of the main text:
\begin{equation}
    \Delta\Omega^{\dag} = \Delta P V^* + \gamma_\mathrm{lv} (A^*_\mathrm{lv} +\cos\theta_Y A_\mathrm{sv}^*)+ \tau l^*_\mathrm{slv}
    \label{nucleation_theory}
\end{equation}
where $V^*$, $A^*_\mathrm{lv}$, $A^*_\mathrm{sv}$, and $l^*_\mathrm{slv}$ are the volume, liquid-vapor area, solid-vapor area, and triple line length of the critical bubble, respectively; $\tau$ is the line tension, and $\Delta P$ can be approximated with the liquid pressure alone $P_\mathrm{l}$. We can assume that the (average) time for the nucleation of the bubble is an activated process described by an Arrhenius kinetics:
\begin{equation}
    \frac{1}{t} \sim \frac{1}{t_\mathrm{sp}} \exp{(-\Delta \Omega^\dag /k_B T)}.
    \label{arrhenius}
\end{equation}
Here, $1/t_\mathrm{sp}$ is a prefactor and $1/t$ is the rate at which nucleation is observed. From Eqs.~\eqref{nucleation_theory} and \eqref{arrhenius}, one can obtain a relation between the rate of nucleation and the liquid pressure at which extrusion occurs:
\begin{equation}
    P_\mathrm{ext} = \frac{k_B T}{V^\ast} \ln{\left(\frac{t}{t_\mathrm{sp}} \right)} - \frac{\gamma_\mathrm{lv} (A^*_\mathrm{lv} +\cos\theta_Y A_\mathrm{sv}^*)+ \tau l^*_\mathrm{slv}}{V^*}.
    \label{pext}
\end{equation}
From equation \eqref{arrhenius}, we can see that having $t=t_\mathrm{sp}$ is equivalent to suppress the barrier for nucleation, which leads to identify the second term in Eq.~\eqref{pext} as the spinodal extrusion pressure $P_\mathrm{ext,sp}$. 
According to Ref.~\cite{guillemot2012}, the prefactor can be written as $1/t_\mathrm{sp}=(L \nu)/b$, where $L$ is the pore length and $\nu$ and $b$ are a microscopic frequency and length scale, respectively. In experiments and simulations $L$ is different, while the reference quantities $\nu$ and $b$ should be comparable, such that $t_\mathrm{sp}^\mathrm{exp}/t_\mathrm{sp}^\mathrm{sim}=L_\mathrm{sim}/L_\mathrm{exp}$. Equation~\eqref{pext} may be applied to an experiment in which the applied pressure $P_\mathrm{ext}^\mathrm{exp}$ corresponds to a nucleation rate $1/t^\mathrm{exp}$ and $t_\mathrm{sp}^\mathrm{exp}=(L^\mathrm{exp}\nu)/b$, or to a simulation in which the applied pressure $P_\mathrm{ext}^\mathrm{sim}$ is the pressure at which one observes nucleation within the duration of the constant-pressure simulation window, $t^\mathrm{sim}=10$ ns, and $t_\mathrm{sp}^\mathrm{sim}=(L^\mathrm{sim}\nu)/b$. The difference between the experimental and simulated extrusion pressures is thus:
\begin{equation}
    P_\mathrm{ext}^\mathrm{exp}-P_\mathrm{ext}^\mathrm{sim}= \frac{k_B T}{V^\ast} \ln{\left(\frac{t_\mathrm{exp}L_\mathrm{exp}}{t_\mathrm{sim}L_\mathrm{sim}} \right)}
    \text{ ,}
\end{equation}

which holds because both the spinodal pressure and the extrusion critical volume are expected to be close in experimental and simulation conditions, as they do not depend neither on $t$ nor on $L$.
The estimated difference between experimental and simulated extrusion pressures, which takes into account the different pore lengths and timescales, is of the order of 13 MPa for long pores ($L^\mathrm{exp}=10\,\mu$m) and for $t^\mathrm{exp}=1$~s; being a systematic error, it can be corrected for.}

{\subsection*{Intrusion pressure}}

{Within the classical capillarity framework, the coexistence pressure is defined as the pressure at which the grand potential of the empty and filled pore are equal \cite{giacomello2020}. Equating the grand potentials $\Omega$ of the filled and empty states one gets:
\begin{equation}
    P_\mathrm{coex} = {-\frac{2\gamma_\mathrm{lv}\cos\theta_Y}{R}} -\frac{2\gamma_\mathrm{lv}}{L}  - \frac{4\tau_\mathrm{lv}}{RL} 
    \label{eq:coex}
\end{equation}
The coexistence pressure is thus given by a first term representing the spinodal intrusion pressure $P_\mathrm{int,sp}$ as predicted by Laplace's law, and two additional terms connected with the finite length of the pore. }

{In the experimental case, the typical length of the nanopores is such that the second and third terms are negligible. As an example, Ref.~\cite{guillemot2012} reports $L=10\,\mu$m and $R=1.34$~nm for MCM-41. The experimentally measured intrusion pressure thus is:
\begin{equation}
    P_\mathrm{int}^\mathrm{exp} \simeq P_\mathrm{coex} \simeq P_\mathrm{int,sp} = {-\frac{2\gamma_\mathrm{lv}\cos\theta_Y}{R}} 
\end{equation}
This \emph{long pore limit} of experimental relevance is represented in Fig.~\ref{G_sketch} by  red dashed lines. These lines correspond to neglecting the free energy contributions from the menisci; this approximation causes the free energy profile to be flat at $P_\mathrm{int}^\mathrm{exp} \simeq P_\mathrm{coex}\simeq P_\mathrm{int,sp}$. In other words, for long pores not only the two metastable states have the same free energy, but also there is no free energy barrier separating them.}

{For intrusion, our simulations have two main differences from experimental pores: the pores are considerably shorter in length and the simulated time is much shorter, $L=20$~nm and $t\approx10$~ns. In this case, the simulated intrusion pressure must fall between the coexistence and spinodal pressure in Eq.~\eqref{eq:coex}, $P_\mathrm{coex}<P_\mathrm{int}^\mathrm{sim}<P_\mathrm{int,sp}$. One could thus assess the maximum deviation from the experimental value $P_\mathrm{int}^\mathrm{sim}-P_\mathrm{int}^\mathrm{exp}= P_\mathrm{coex}-P_\mathrm{int,sp}=-2\gamma_\mathrm{lv}/L  - 4\tau_\mathrm{lv}/(RL)  \approx -7.2 $~MPa. This is a systematic error which can be accounted for. We remark that this pessimistic estimate would only hold if each constant-pressure simulation had an infinite duration, long enough to assess the equal probability of occupation of the filled and empty pore, i.e., to identify $P_\mathrm{coex}$ in Eq.~\eqref{eq:coex}. This is because, in simulations, the free energy profile is not flat and the two minima are separated by a free-energy barrier due to the presence of the menisci (black lines in Fig.~\ref{G_sketch}). In contrast to this infinite duration limit, actual simulation windows are short (10 ns). So, intrusion is observed in simulations only when the intrusion barrier due to the presence of the menisci is almost completely suppressed by an increased pressure, i.e., close to the spinodal pressure. In actual simulations, the intrusion pressure is closer to the spinodal intrusion pressure, i.e., to the experimental one, than to the coexistence one. Using the same functional form of Eq.~\eqref{pext}, we can estimate more accurately how far from the spinodal and thus from the experimental value the actual simulation pressure is, which results in $P_\mathrm{int}^\mathrm{sim}-P_\mathrm{int}^\mathrm{exp}<50$~kPa, having used as the intrusion critical volume $V^\ast$ the entire pore volume in the simulations, $t=10$~ns ,and $t_\mathrm{sp}=0.5$~ns. Simulated intrusion pressures should thus be in excellent agreement with experimental ones, once the correct surface tension is considered.}

    \begin{figure}[ht!]
    \centering
      \includegraphics[width=0.7\linewidth]{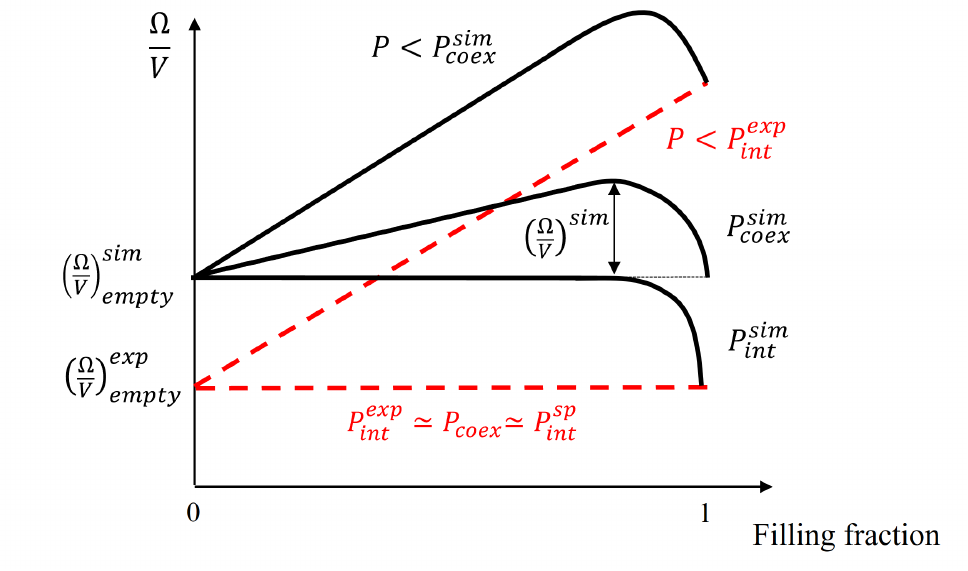}
      \caption{{Comparison between the trends of the free energy profiles at different pressures. Red dashed lines refer to the experimental \emph{long pore limit}, in which it is possible to neglect the free energy contributions from the menisci. Black lines are relative to the small pores, relevant to our simulations, in which the menisci contributions to the free energy are kept into account.}} 
      \label{G_sketch}
    \end{figure}

{\section*{Effect of silica hydrophilicity}
\begin{figure}[ht!]
    \centering
      \includegraphics[width=0.9\linewidth]{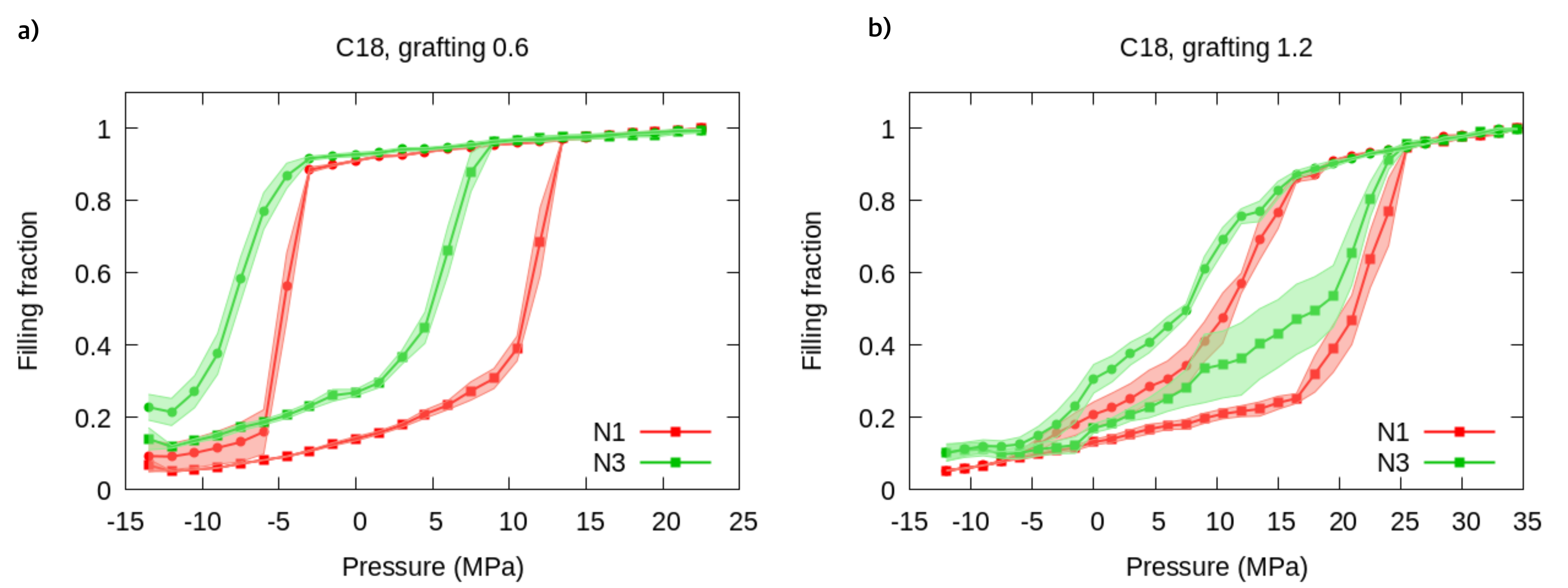}
      \caption{{Comparisons betweeen I/E cycles with different hydrophobicity of the silica surfaces for C18-0.6 (a) and C18-1.2 (b) systems. Red curves refer to silica surfaces modeled with N1 beads, corresponding to a water contact angle of 75°. Green curves refer to more hydrophilic silica surfaces, modeled with N3 beads, corresponding to a water contact angle of 46$^\circ$.}} 
      \label{silica_N3}
    \end{figure}
As the experimental data on the contact angle of water on silica suggest that the contact angle is sensitive to the details of the surface preparation, we performed a set of five independent I/E cycles using a model silica with a contact angle of $46^{\circ}$ (corresponding to bead N3 in the Martini 3 model). The results are shown in Fig.~\ref{silica_N3} for C18 at the extreme grafting densities of 0.6 and 1.2 grps/nm$^2$. Coherently with what we observed for the less hydrophilic silica, a larger exposure of hydrophilic surface at low density leads to a left shift of the I/E cycles. At high density, the effect is less pronounced, as the silanes prevent water from direct interaction with the silica surface.
}

\begin{figure}[ht!]
\centering
  \includegraphics[width=0.7\linewidth]{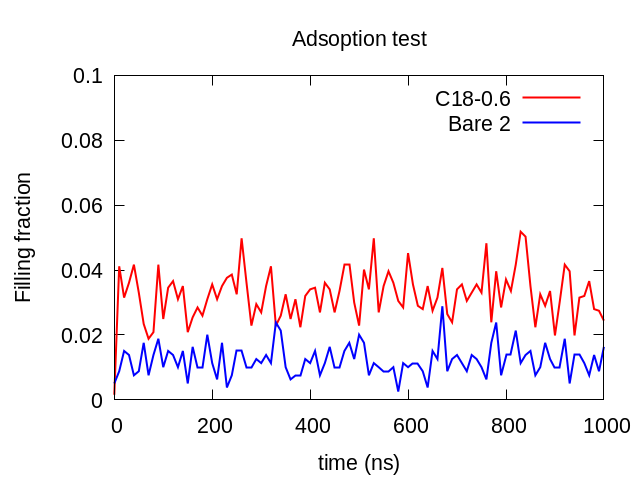}
  \caption{{ Filling fraction as a function of time, reporting results for a long (1 $\mu \text{s}$) simulation at atmospheric pressure for the \emph{bare} hydrohpilic (Bare 2) and grafted (C18-0.6) nanopore cases. Absence of any significant adsorption from the vapor phase at fixed liquid pressures below $P_{int}$ was observed for both cases.
  }} 
  \label{ads}
    \end{figure}

\clearpage

\bibliography{bib}

\clearpage